\documentclass[twocolumn]{aastex631}

\newcommand{\sss}{\scriptscriptstyle}
\newtheorem{lemma}{Lemma}[section]

\usepackage[ruled,vlined]{algorithm2e}
\usepackage{mathrsfs}
\usepackage{subfigure}
\usepackage{amsmath}
\usepackage{bm}

\begin{document}
	
	\title{The spherical Fast Multipole Method (sFMM) for Gravitational Lensing Simulation}
	
	
	\author{Xingpao Suo}
	\affiliation{Institute for Astronomy, School of Physics, Zhejiang University,  Hangzhou 310027, China}
	
	\correspondingauthor{Xingpao Suo}
	\email{xpsuo@zju.edu.cn}
	
	\author{Xi Kang}
	\affiliation{Institute for Astronomy, School of Physics, Zhejiang University,  Hangzhou 310027, China}
	\affiliation{Purple Mountain Observatory, Nanjing 210008, China}
	\email{kangxi@zju.edu.cn}
	
	\author{Chengliang Wei}
	\affiliation{Purple Mountain Observatory, Nanjing 210008, China}
	
	\author{Guoliang Li}
	\affiliation{Purple Mountain Observatory, Nanjing 210008, China}

	\begin{abstract}
		In this paper, we present a spherical Fast Multipole Method (sFMM) for ray tracing simulation of gravitational lensing (GL) on a curved sky.
		The sFMM is a non-trivial extension of the Fast Multiple Method (FMM) to sphere $\mathbb S^2$, and it can accurately solve the Poisson equation with time complexity of $O(N)\log(N)$, where $N$ is the number of particles. 
		It is found that the time complexity of the sFMM is near $O(N)$ and the computational accuracy can reach $10^{-10}$ in our test.
		In addition, compared with the Fast Spherical Harmonic Transform (FSHT), the sFMM is not only faster but more accurate, as it has the ability to reserve high-frequency components of the density field.
		These merits make the sFMM an optimum method to simulate the gravitational lensing on a curved sky, which is the case for upcoming large-area sky surveys, such as the Vera Rubin Observatory and the China Space Station Telescope.
	\end{abstract}

	\section{Introduction}
	Gravitational lensing (GL) is a powerful probe to detect the distribution of dark matter \citep[e.g.,][]{hoekstra2004properties}, constrain the cosmological parameters \citep[e.g.,][]{KiDs}, and search for extra-solar planets \citep[e.g.,][]{dominik2002planet}. These merits make GL a primary goal for on-going or upcoming large-sky surveys, such as the Kilo Degree Survey (KiDS\footnote{\url{http://kids.strw.leidenuniv.nl}}),
	the Dark Energy Survey (DES\footnote{\url{https://www.darkenergysurvey.org}}),
	the Vera Rubin Observatory, and the China Space Station Telescope (CSST).
	
	To fully understand the systematic errors in observations of GL, simulation with accurate predictions of GL effect is very essential \citep[e.g.,][]{takahashi2017full}. The ray-tracing technique has been widely developed and used in the simulation of GL effect \citep[e.g.,][]{jain2000ray, taruya2002lognormal, vale2003simulating, hilbert2009ray, teyssier2009full, hezaveh2011effects,becker2013calclens}.

	In order to implement an N-body-based GL simulation by ray-tracing, the Poisson equation (PE\footnote{Hereafter, PE refers to the Poisson equation with point or point-like sources, unless we specifically declare.}) should be solved in 3-D space to get the deflection angle of the light beam \citep[see, e.g.,][]{breton2021magrathea}, and it is well known that solving PE involves huge numerical complexity. 
	To simplify the calculation, one introduces Born, flat-sky, and multi-planes approximations. With them, the deflection angle of light could be obtained by projecting all the mass in the space onto a few of 2-D planes and solving the 2-D PE for each plane \cite[e.g.,][]{amara2006simulations, hilbert2009ray}.
	This method can greatly improve computational efficiency,
	but it is only appropriate for small-scale sky surveys. To meet the demand of future large-scale sky surveys, a spherical ray-tracing technique has been developed by projecting the mass onto the sphere instead of the plane and solving the spherical PE to obtain the deflection angle \citep[e.g.,][]{becker2013calclens}, and this method has been used to produce a full-sky ray-tracing of GL with a large-volume N-body simulation \citep{wei2018full}.

	Since solving the PE is one of the most key and time-consuming parts in ray-tracing, a fast and accurate PE solver is of crucial importance. There are three kinds of main algorithms to solve the PE \citep[see][]{jain2000ray}:\newline
	
	1. \textbf{Directly sum} \citep[e.g.,][]{moller1998strong,moller2001strong,giocoli2017fast,legin2021simulation,sonnenfeld2021statistical}. In this method, one directly sums the potential generated by some sources on the \textit{target points}\footnote{Throughout this paper, \textit{target points} refers to the points on which one calculates the lensing quantities such as lensing potential and deflection angle.}.
	Suppose we have $N$ sources and $M$ target points, then the calculation complexity of this method will be $O (MN)$. 
	
	2. \textbf{Tree Code} \citep[e.g., ][]{wambsganss1998testing,wambsganss1999gravitational,thompson2010teraflop,10.1093/mnras/stu1859}. In this method, one assigns all the particles into a self-adaptive and hierarchical quad-tree, then treats all the particles in the same subtree as a pseudo-lens. Because a bunch of particles can be treated as a single pseudo-lens, this method accelerates the calculation. The time complexity of it is $O (M\log (N))$ \citep{wambsganss1999gravitational}. This method is similar to the Fast Multipole Method (FMM), however, computational redundancies still exist in it. Besides, the previous work on this method has not developed a method suitable for simulating GL on a sphere.
	
	3. \textbf{FFT or FFT-based method} \citep[e.g.,][]{amara2006simulations, meneghetti2008realistic, hilbert2009ray,xu2021accurate}. In this method, one assigns the particles onto meshes, then applies FFT to transform the meshed density into Fourier space, in which the PE will be merely multiplication. Finally, one applies inverse FFT to get the potential in real space.  On a sphere, the FFT-based Fast Spherical Harmonics Transform \citep[FSHT, see][]{press2007numerical} can be used, but it has a few defects. Firstly, the complexity of the FSHT is still time-consuming as $O(N_{grid}^{3/2})$ where $N_{grid}$ is the number of mesh grids. Secondly, like FFT, the FSHT requires one to uniformly sample the field on the isolatitude lines, and the information beyond Nyquist frequency $\omega_{N}$ will be lost during the sampling. So to obtain higher frequency components of a small high-density region, one has to increase the resolution globally, followed by amounts of additional calculation. Lastly, the associated Legendre functions used in the FSHT can only be generated via slow recursions \citep[see][]{gorski2005healpix, press2007numerical}. Moreover, the computational stability will decrease with increasing order of the associated Legendre functions.  These drawbacks not only increase the time complexity of the FSHT, but also limit the FSHT's ability to obtain high frequency information. \newline

	To overcome the shortcoming of the above methods, we have developed a method, named the spherical Fast Multipole Method (sFMM) to solve the PE on a sphere $\mathbb S^2$ for large or full-sky GL ray-tracing based on N-body simulation.
	As one of the most successful algorithms, the standard FMM has been originally presented by \cite{greengard1987fast} to solve the PE in 2-D flat space.
	With the time complexity of $O(N)$, it can calculate the potential up to the machine precision. Moreover, the high frequency information is preserved since it occupies a hierarchical data structure and calculates the potential in different length scales.
	Because of these advantages, the standard FMM has been expanded to a 3-D flat space version \citep{cheng1999fast}, a continuous source version \citep{ethridge2001new}, a kernel-independent version \citep{ying2004kernel}, and so on.
	Served as another non-trivial expansion of the standard FMM, the sFMM proposed in this paper also inherits its advantages, as we will show in Section \ref{SOLVE} and \ref{TEST}.
	In addition, we use a self-adaptive and hierarchical data structure in the implementation, which enhances the ability of the sFMM to deal with inhomogeneous mass distribution. With these merits, we hope it will be an optimized method in GL ray-tracing.
	
	This paper is structured as the follows. In Section \ref{THEORY} we clarify the conventions used in the paper and briefly give the formulae in spherical GL simulation. In Section \ref{SOLVE} we introduce the sFMM and show how it solves the PE in a fast and accurate way. In Sections \ref{TEST} and \ref{CONCLUSION}, we test our method and give the conclusion, respectively. 
	

	\section{Spherical GL Ray-tracing}
	\label{THEORY}
	In this section, we review some basic concepts and briefly derive some formulae of spherical GL ray-tracing. Particularly, the Green function of spherical Laplacian is given as the kernel of the sFMM.
	
	\subsection{The Deflection of Light}
	Following the idea of general relativity, the deflection of light is the consequence of a curved time-space. So we begin with a slightly perturbed Friedmann-Robertson-Walker metric of the universe
	\begin{align}
		ds^2 = a^2 (\tau)\left[ \left (1 + \frac{2\Phi }{c^2}\right) c^2 d\tau^2 - \left (1-\frac{2\Phi }{c^2}\right) \left ( dr^2 + f^2_K (r) d\Omega^2\right)\right]\ , \label{FRW_MET}
	\end{align}
	where $\tau$ is the conformal time, $r$ is the comoving radial distance, $d\Omega^2 \equiv  d\theta^2 + \sin^2\theta d\phi^2$, $a=1/ (1+z)$ is the scale factor normalized to unity today, $\Phi= \Phi (\tau, r,\theta,\phi) $ is the Newtonian peculiar gravitational potential, and in the universe with space-curvature $K$, the comoving angular diameter distance is given by
	\begin{align}
		f_K (r) = \left\{
		\begin{matrix}
			&\frac{1}{\sqrt{K}}\sin{\sqrt{K}r}    & K>0\\
			&r   & K=0\\
			&\frac{1}{\sqrt{-K}}\sinh{\sqrt{-K}r}  &  K<0\ .
		\end{matrix}
		\right. \label{F_K}
	\end{align}
	In a weakly perturbed universe,  by solving the geodesic equation of light, the relation between  angular position $\bm \theta$ of a source at comoving distance $r_s$ and the observed image position $\bm \theta_0$  of the source on the sky is given by \citep[see, e.g.,][]{jain1997cosmological}
	\begin{align}
		\bm\theta (\bm \theta_0, r_s)
		&= \bm \theta_0 - \frac{2}{c^2} \int_0^{r_s} dr \frac{f_K (r_s - r)}{f_K (r_s) }\bm\nabla_\perp \Phi\left ( \bm\theta (\bm\theta_0, r),r\right)\label{BASE_EQU}\\
		&=\bm\theta_0 - \int_0^{r_s} \frac{f_K (r_s - r)}{f_K (r_s) } d\bm\alpha (\bm\theta (\bm\theta_0,r), r )\ ,\label{BASE_EUA2}
	\end{align}
	where $\bm\nabla_\perp$ denotes the co-variant transverse comoving gradient operator and in the layer of comoving distance from $r$ to $ r + dr $, the deflection angle is given by
	\begin{align}
		d\bm \alpha (\bm\theta, r) =  dr \frac{2}{c^2}\bm\nabla_\perp \Phi\label{DEF_ALP}\left ( \bm\theta ,r\right)\ .
	\end{align}
	
	The main difficulty to solve the integral Equation  (\ref{BASE_EQU}) numerically comes from the gravitational potential. Strictly speaking, we need to solve the PE in the whole 3-D space. But it's almost impossible if the space we consider is extremely large. To simplify this process, one  introduces the method below \citep[see, e.g.,][]{jain2000ray,becker2013calclens}.
	
	\begin{figure*}
		\centering
		\includegraphics[width=.618\linewidth]{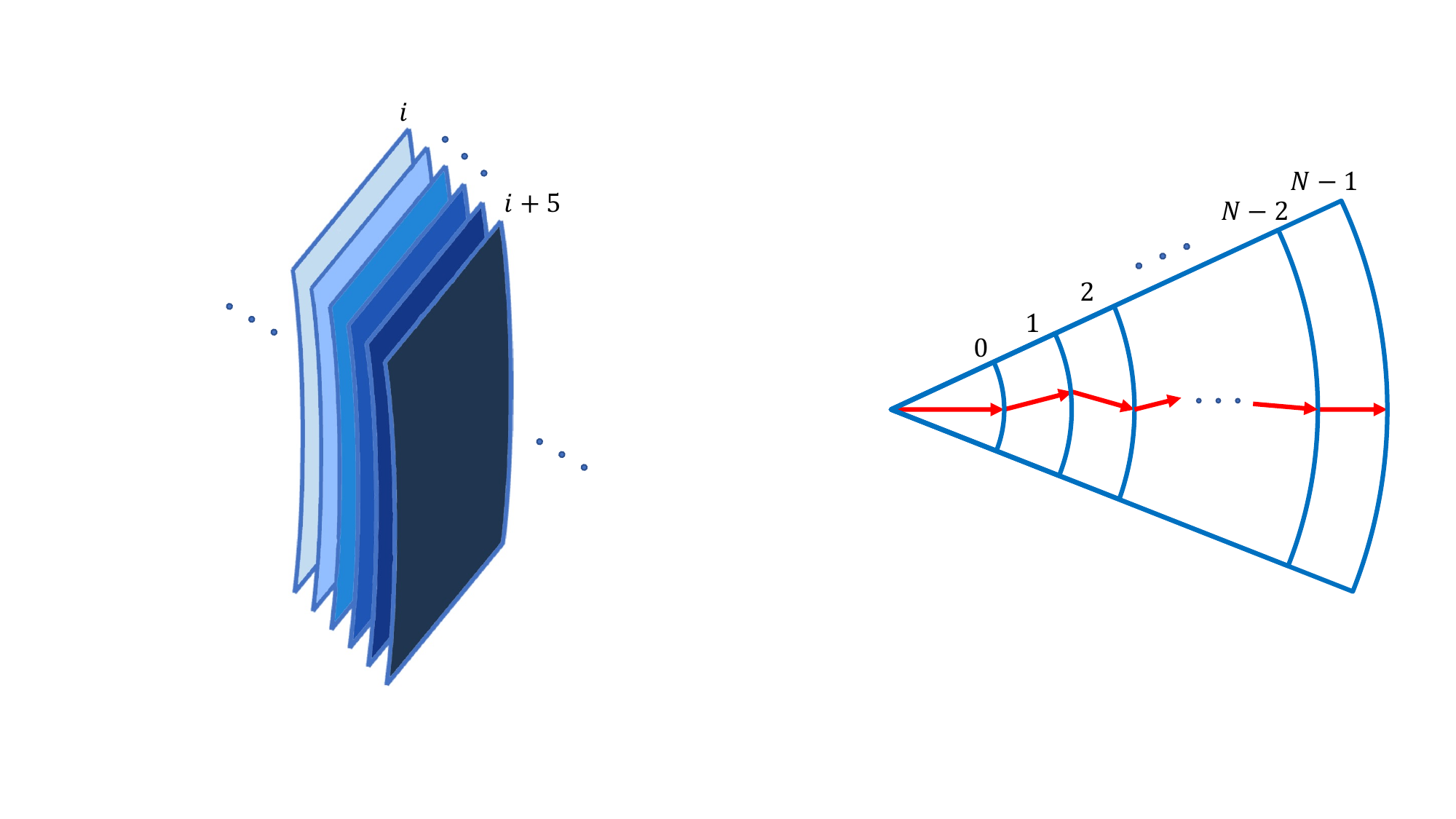}
		\caption{The multi-layers in GL simulation. The left and right panels show the 3-D view and longitudinal cross-section of multi-layers for the spherical ray-tracing, respectively. In the right panel, we show the propagation of the light ray with the backtracking method. }
		\label{layer}
	\end{figure*}
	Firstly, one divides the space into $N_L$ layers perpendicular to the line-of-sight. The comoving radius of the $i$-th layer is from $r_{i-\frac{1}{2}}$ to $r_{i+\frac{1}{2}}$. Then one applies the backtracking method  (i.e., ejecting a light ray from the observer back to the source) as depicted in the right panel of Figure \ref{layer}.
	Labeling the angular position of a photon as $\bm \theta_i$ when it just entered the $i$-th layer, the discretization of Equation  (\ref{BASE_EUA2}) gives
	\begin{align}
		&\bm\theta_{i+1} =  \bm \theta_{i} - \frac{f_K (r_s - r_i)}{f_K (r_s) } \bm \alpha_i(\bm \theta_i)\ ,\label{DIS_PRO}
	\end{align}
	where $i=0,1,..., n_{\sss L}-1$ is the layer's number, $r_i$ is the comoving distance in the middle of the $i$-th layer,
	$\bm\alpha_i \equiv \int_{r_{i-\frac{1}{2}}}^{r_{i + \frac{1}{2}}} d\bm\alpha$
	is the deflection angle in the $i$-th layer, $n_{\sss L} = 0, 1, ..., N_L-1$, and in the $n_{\sss L}$-th layer, the light beam meets the source. With two approximations, one obtains (see Appendix \ref{APPENDIX:A})
	\begin{align}
		\bm \alpha_i (\bm\theta_i)  = \bm\nabla \psi_i  (\bm\theta_i) \label{ALP_PSI}
	\end{align}
	with
	\begin{align}
		\psi_i (\bm\theta) = \int d^2 \theta' \kappa_i (\bm \theta') \mathcal G (\bm\theta,\bm\theta')\ ,\label{PSI_DEF}
	\end{align}
	where $\bm\nabla$ is the co-variant gradient operator on the sphere,  $\mathcal G (\bm\theta,\bm\theta') $ is the Green function, $\psi_i (\bm\theta)$ and $\kappa_i (\bm\theta)$ are the lens potential and the dimensionless surface density  of the $i$-th layer, respectively. For the convenience of numerical work, our definitions for $\kappa$ and $\mathcal G$  are twice over and half of the conventional ones, respectively.  In spherical coordinate system, 
	\begin{align}
		\mathcal G (\bm\theta, \bm\theta')
		&= \frac{1}{2\pi} \log\sin\left(\frac{\Theta}{2}\right) \nonumber \\
		& = \frac{1}{4\pi} \left[ \log \left(1 - \bm e_r (\theta,\phi) \cdot \bm e_r (\theta', \phi')\right) -\log 2 \right]\ ,\label{GRE_FUN}
	\end{align}
	where $\Theta = \arccos{\left(\bm e_r(\theta,\phi) \cdot \bm e_r(\theta',\phi')\right)}$ is the angular distance between $(\theta, \phi)$ and $(\theta', \phi')$.
	Obviously, if $\Theta \to 0$, $\mathcal G(\bm \theta, \bm\theta')$ degenerates to the Green function of 2-D plane. 
	
	The dimensionless surface density $\kappa_i$ can be constructed from the $i$-th layer's column density $\Sigma_i (\theta,\phi)$ with \citep[see also][]{becker2013calclens}
	\begin{align}
		\kappa_i (\theta,\phi) &
		= \frac{8\pi G}{c^2a (r_i)}f_K (r_i) \Sigma_i (\theta,\phi)\ ,\label{KAP_DEF}
	\end{align}
	where $a (r_i)$ is the scale factor when the light ray is in the radial distance $r_i$.
	Note that Equation  (\ref{PSI_DEF}) is equivalent to
	\begin{align}
		\Delta \psi_i = \kappa_i - \bar \kappa_i\ , \label{POI_EQU}\
	\end{align}
	where $\Delta $ is the spherical co-variant Laplacian, $\bar\kappa_i \equiv 1/(4\pi) \int_{\mathbb{S}^2} d^2 \theta' \kappa(\theta') $ is the background density or the average density over the entire sphere of the $i$-th layer.   This means that the Laplacian of $\psi_i$ only relates to the overdensity.   Meanwhile, one should note that the background density will contribute a constant to $\psi_i$ itself. Although a constant in the lensing potential has no effect on the deflection angle, its value is to be set appropriately. In our implement, it is determined to ensure the  lensing potential
	to fit
	\begin{align}
		\int_{\mathbb S^2} d^2  \theta \psi_i(\theta) = 0\ .\label{STAND_CONDITION}
	\end{align}
	In this way, the  potential is consistent with the potential from the FSHT. 

	Solving the PE  (\ref{POI_EQU}) or equivalently calculating Equation  (\ref{PSI_DEF}) is time-consuming because to get the value of the lensing potential field on a target point, one needs to sum the contributions from all particles over the considered region.

	\subsection{The Distortion of Galaxy Image}
	In many cases, the deflection angle of light can't be measured since one can not obtain the real position of a source. Hence, we need to find some other observable quantities. It's natural to study the Hessian matrix of lensing potential, which gives the information on the distortion of the galaxy image in GL, and it can be measured by, for example, the statistic of galaxies' ellipticity.
	
	With the normalized basis $\bm e_\theta(\theta,\phi)$ and $\bm e_\phi(\theta,\phi)$, the Hessian matrix $\bm H$ of lensing potential $\psi$ is given by \citep[see][]{de2010halo}
	\begin{align}
		\bm H =  \bm\nabla^2 \psi
		= \left( 
		\begin{matrix}
			\partial_\theta^2\psi &
			\partial_\theta \left ( \frac{1}{\sin\theta} \partial_\phi \psi \right) \\
			\partial_\theta \left ( \frac{1}{\sin\theta} \partial_\phi \psi \right) & 
			\frac{1}{\sin^2\theta} \partial_\phi^2 \psi + \cot (\theta) \partial_\theta \psi
		\end{matrix}
		\right)\ .
	\end{align}
	Note that the Laplacian $\Delta = \mathrm{tr}\bm \nabla^2 $, so we can decompose $\bm H$ as trace part $ (\kappa - \bar\kappa)/2$  and trace-less part $\bm\gamma$ as in the 2-D plane case
	\begin{align}
		\bm H &=\frac{1}{2} \mathrm{tr} (\bm H) \bm I + \bm\gamma \nonumber \\
		& = \frac{1}{2}\left ( \kappa - \bar  \kappa \right) \bm I  + \left ( 
		\begin{matrix}
			\gamma_1 &\gamma_2\\
			\gamma_2 & -\gamma_1
		\end{matrix}
		\right)\ , \label{GAM_DEF}
	\end{align}
	where in the last step, Equation  (\ref{POI_EQU}) is used, and we define the components of shear  in bases of $\bm e_\theta$ and $\bm e_\phi$  as
	\begin{align}
		&\gamma_1 \equiv \frac{1}{2}\left(\bm H_{11} - \bm H_{22}\right)\ {\rm and } \\
		&\gamma_2 \equiv \bm H_{12}=\bm H_{21}\ ,
	\end{align}
	respectively.
	Finally, the magnification of a small source is given by 
	\begin{align}
		\mu = \frac{1}{\det{(\bm I-\bm H)}} = \frac{1}{ (1- (\kappa-\bar \kappa)/2)^2 - \gamma_1^2 - \gamma_2^2}\ .
	\end{align}
	
	These equations are just for one lensing plane. Taking the number of layers  $N_L>1$, the Hessian can be propagated from one to another plane \citep[see][]{hilbert2009ray,becker2013calclens}. 

	\section{Solving the PE on a Sphere: the sFMM}
	\label{SOLVE}
	In this section, we demonstrate the details of the sFMM. As an extension of FMM, the sFMM uses the multipole expansion on a sphere to approximate the far field and employs the so-called \textit{up-pass} and \textit{down-pass} procedures to reduce the redundancy in the calculation. 
	\subsection{Basic Idea\label{BasicIdea}}
	In the GL ray-tracing based on N-body particles, the lensing potential is generated by point sources, which means one can write Equation (\ref{PSI_DEF}) as 
	\begin{align}
		\psi (\bm\theta)  = \sum_{j=0}^{N-1}q_j \mathcal G (\bm\theta, \bm\theta_j)\ ,\label{SUM}
	\end{align}
	where $q_j$ denotes mass of the $j$-th particle, following the conventions in \cite{greengard1987fast} and $\bm \theta_j$ is the $j$-th particle's position.
	To calculates the summation in Equation (\ref{SUM}) with an adaptive and hierarchical way, we first divide the sphere into a tree structure according to the particles' distribution. The division is easy for a flat space, but non-trivial for a sphere. Giving $m$, the maximal number of particles in leaf boxes, we use HEALPix\footnote{\url{https://healpix.sourceforge.io}}--the Hierarchical Equal Area isoLatitude Pixelization \citep{gorski2005healpix} data structure to perform the division with the following steps:
	\begin{enumerate}
		\item  Divide the sphere into 12 \textit{base boxes} and assign all the particles into these boxes.
		\item For each leaf box in the tree, if it contains more than $m$ particles, divide it into 4 child boxes, then assign the particles in it into its child boxes.
		\item Repeat Step 2 until there is no box containing more than $m$ particles in the tree.
	\end{enumerate}
	\begin{figure*}
		\centering
		\includegraphics[width=.618\linewidth]{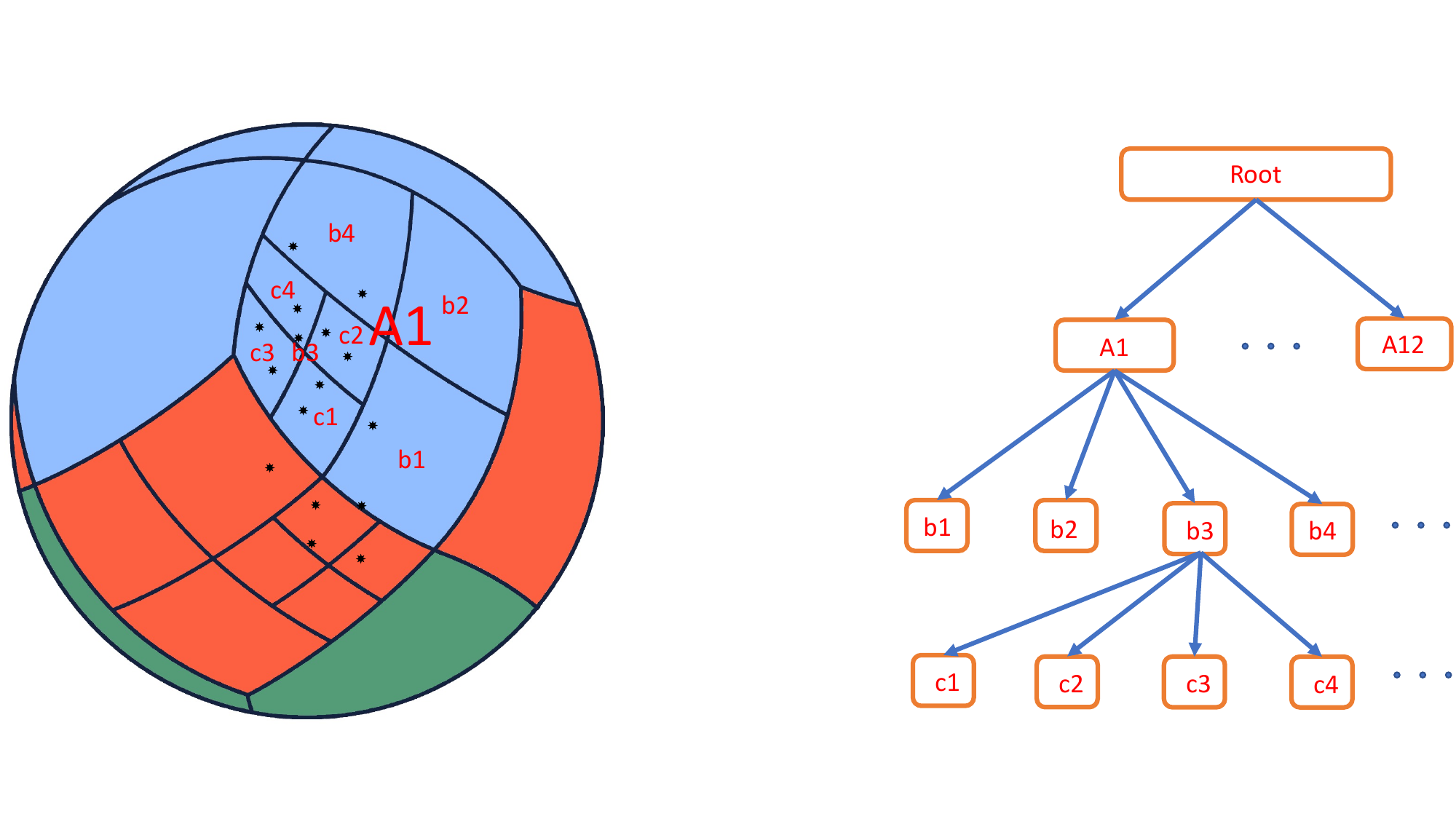}
		\caption{An example of data structure we use in the sFMM. The left panel shows the particles on the sphere and an adaptive, hierarchical division according to these particles with HEALPix. The whole sphere is divided into 12 boxes firstly, then the boxes containing more than $m=2$ particles are divided into four boxes recursively until each box contains no more than $m$ particles. The right panel shows the corresponding data structure, in which the root node has 12 child boxes and each of them is a quad-tree. }
		\label{quadtree}
	\end{figure*}
	
	Note that in practice, the division will stop if the depth of the tree reaches the \textit{maximum depth} $r_{max}$. Beyond $r_{max}$, two particles in the same box are so close that they can be merged. In our code, we set $r_{max} = 29$ by default, which is  enough for most tasks (corresponding box radius $\approx 4\times 10^{-4}\mathrm{arcsec}$).
	
	After the division is finished, we will get a quasi-quad-tree, as depicted in Figure \ref{quadtree}. 
	Then, taking the center of a box $\mathcal B$ as the origin, we can build a local spherical coordinate system and use a multipole expansion defined in it to approximate the far potential generated by the particles in $\mathcal B$. 
	
	Giving $P$ with its coordinate $(\theta_P,\phi_P)$, the spherical coordinate $K_P$ with $P$ as the origin will take $\bm e_r(\theta_P,\phi_P) $ and $\bm e_\theta(\theta_P, \phi_P)$  as the \textbf{zenith} and \textbf{azimuth} references, respectively. Furthermore, the coordinate in $K_P$ is denoted as $(\theta^{\sss (P)}, \phi^{\sss (P)})$. Note that throughout this paper, we keep these conventions.
	
	A lemma can be found to construct the multipole expansion in the local coordinate system:
	\begin{lemma}\label{LEMMA1}
		(Multipole expansion on the sphere) In the local coordinate system $K_S$, suppose that $n$ point sources of masses \{$q_i$, $i = 1,2,...,n$\} are located at points \{$(\theta_i^{\sss (S)},\phi_i^{\sss (S)})$, $i=1,2,...,n$ \} with $\theta_i^{\sss (S)} < \theta_r^{\sss (S)}$. Then for any point $ (\theta^{\sss (S)},\phi^{\sss (S)})$ with $\theta^{\sss (S)}>  \theta_r^{\sss (S)}$, the potential $\psi$ generated by these sources can be written as
		\begin{align}
			\psi (\theta^{\sss (S)},\phi^{\sss (S)}) & = C + Q \log (1 - \cos\theta^{\sss (S)}) \nonumber \\
			&	+ 2 \Re\left (\sum_{l=1}^\infty \frac{A_l}{z^l (\theta^{\sss (S)},\phi^{\sss (S)})}\right) \label{LEMMA1:EXP}
		\end{align}
		that converges for $\theta^{\sss (S)} > \theta^{\sss (S)}_r$,
		where $z^l(\theta,\phi) \equiv \tan^l \left ({\theta}/{2}\right) e^{i l\phi} $ is the base function of spherical multipole expansion,
		\begin{align}
			&C = \frac{1}{4\pi}\sum_{i=1}^n 2 q_i \log\cos \frac{\theta^{\sss (S)}_i}{2} - Q  \log 2\nonumber \ , \\
			&Q = \frac{1}{4\pi}\sum_{i=1}^n q_i\nonumber \ , \ \text{and}\\
			&A_l = - \frac{1}{4\pi } \frac{1}{l}\sum_{i=1}^n q_i z^l (\theta^{\sss (S)}_i,\phi^{\sss (S)}_i)\ .
		\end{align}
	\end{lemma}
	
	One can prove Lemma \ref{LEMMA1} and following lemmas with the hints in Appendix \ref{APP_B}. Besides, the truncation error bound of Expansion  (\ref{LEMMA1:EXP}) is suppressed as
	\begin{align}
		e_p\equiv & \left|  \psi - C - Q \log (1 - \cos\theta^{\sss (S)}) - 2 \Re\left (
		\sum_{l=1}^p \frac{A_l}{z^l (\theta^{\sss (S)},\phi^{\sss (S)})}
		\right)   \right| \nonumber \\
		& \leq \frac{\tilde{Q}}{2\pi} \sum_{l = p+1}^{\infty}\frac{1}{l} \left( \frac{\tan(\theta_r^{\sss (S)}/2)}{\tan(\theta^{\sss (S)}/2)}\right)^l \nonumber \\
		&\leq \frac{\tilde{Q}}{2\pi} \frac{1}{p+1} \frac{x^{p+1}}{1-x} \ , \label{MUL_ERR}
	\end{align}
	where
	$\tilde{Q} \equiv \sum_{i=1}^{n}|q_i| $, 
	$x \equiv {\tan(\theta_r^{\sss (S)}/2)} / {\tan(\theta^{\sss (S)}/2)}$, and
	we call $p$ the truncation parameter.
	As an estimation, taking  $x =0.5$, the relative precision of $10^{-16}$  (i.e., the precision of the double float in 64-bits machine) can be achieved by keeping up to $\log_2 (10^{-16}) \approx 53$ multipole items. Obviously, the further the target point from the source box, the more accurate the truncated multipole expansion is.
	Note that hereafter, for a certain $p$, \textit{the point $T$ is 'far' from the source box} means that the Error Bound (\ref{MUL_ERR}) is suppressed to a desired order. In Section (\ref{SFMM:MAC}) we will give a more precise definition of \textit{'far'}.
	
	The multipole expansions of all boxes in the tree can be constructed by employing Lemma \ref{LEMMA1} in corresponding local coordinates.
	After that, one can sum all the potential on a target point $T$ from top (root) to down (leaves). 
	That is, consider each base box in the tree, if $T$ is far from it, sum the potential contribution from it by its multipole expansion, if not, consider its child boxes recursively, until we encounter the leaf boxes near $P$, say, $\mathcal B_L$. 
	The potential contributed from $\mathcal B_L$ can not be calculated from the multipole expansion, so one needs to do it by direct summation. 
	
	Note that in the direct-calculation level, without influencing the calculation of multipole expansions, one can also view each particle as a mass distribution with radial symmetry, which we call the Near Particle Model (NPM). With NPM, the particle will occupy a small space, but can still be viewed as a point outside it. Furthermore, the potential inside the particle can be calculated analytically.  See more details in Section \ref{TEST}. 
	
	One could notice that redundancies exist in the method described above. For example, we need to employ a particle several times to obtain the multipole expansions for boxes in different levels. 
	As in the standard FMM, the sFMM will use two strategies to eliminate these redundancies in the calculation.
	
	\subsection{The Up-pass}
	One strategy comes from the observation that knowing a non-root box $\mathcal B$'s multipole expansion, we can translate it from the center of $\mathcal B$ onto the center of $\mathcal B$'s parent $\mathcal B_P$. The same process can be done for the four children of $\mathcal B_P$, then we get $\mathcal B_P$'s multipole expansion. So firstly, we construct the multipole expansions for all leaf boxes by using Lemma \ref{LEMMA1}, then pass up the expansions from down to top until we encounter the root box. This process is called  \textit{up-pass}.
	To realize it, we have
	
	\begin{lemma}\label{LEMMA2}
		(Translation of a multipole expansion)  Suppose that  in the spherical coordinate system $K_S$,
		\begin{align}
			\psi (\theta^{\sss (S)},\phi^{\sss (S)})  & = C + Q\log (1-\cos (\theta^{\sss (S)}) \nonumber \\
			& +2 \Re \sum_{l=1}^\infty \frac{A_l}{z^l (\theta^{\sss (S)},\phi^{\sss (S)})} \label{LEMMA2:EXP1}
		\end{align}
		is a multipole expansion that converges for  $\theta^{\sss (S)} > \theta^{\sss (S)}_r$,
		where $ (\theta^{\sss (S)},\phi^{\sss (S)})$ the coordinate of a point $P$ in $K_S$,
		then in another spherical coordinate system $K_T$, the Expansion  (\ref{LEMMA2:EXP1}) can be written as
		\begin{align}
			\psi' (\theta^{\sss (T)},\phi^{\sss (T)}) &= C' + Q' \log (1-\cos (\theta^{\sss (T)})) \nonumber \\
			&+2 \Re \sum_{l=1}^\infty \frac{A'_l}{z^l (\theta^{\sss (T)},\phi^{\sss (T)})}\  \label{LEMMA2:EXP2}
		\end{align}
		that converges for $\theta^{\sss (T)} >  \theta_r^{\sss (T)} \equiv  \theta^{\sss (S)}_{\sss T} + \theta^{\sss (S)}_r $,
		where $ (\theta^{\sss (T)},\phi^{\sss (T)})$ the coordinate of $P$ in $K_T$,  
		\begin{align}
			&C' = C  + 2 Q \log\cos\frac{\theta^{\sss (S)}_{\sss T}}{2} + 2 \Re \sum_{l=1}^\infty  (-1)^l e^{- i l \phi^{\sss (S)}_{\sss T}}A_l  \tan^l \frac{\theta^{\sss (S)}_{\sss T}}{2},\nonumber \\
			&Q'=Q\ ,\ \text{and}\nonumber  \\
			&A'_l = \left ( -\frac{Q}{l}\tan^l\frac{\theta^{\sss (S)}_{\sss T}}{2} +  \sum_{l'=1}^\infty  (-1)^{l'} e^{-i l' \phi^{\sss (S)}_{\sss T}} A_{l'} T_{ll'} \right)e^{i l\phi^{\sss (T)}_S}
		\end{align}
		with $ (\theta^{\sss (S)}_{\sss T}, \phi^{\sss (S)}_{\sss T})$ the coordinate of point $T$ (i.e., the origin of $K_T$, similar below) in $K_S$, $ (\theta^{\sss (T)}_{\sss S} = \theta^{\sss (S)}_{\sss T}, \phi^{\sss (T)}_{\sss S})$ the coordinate of point $S$ in $K_T$, and
		\begin{align}
			T_{ll'} \equiv  \sum_{n=1}^{\min (l,l')} C_{l-1}^{n-1}C_{l'}^n\tan^{l+l'-2n}\frac{\theta^{\sss (S)}_{\sss T}}{2} \left (1 + \tan^2 \frac{\theta^{\sss (S)}_{\sss T}}{2}\right)^n\ ,
		\end{align}where $C_n^m$ the binomial coefficients. 
	\end{lemma}
	
	If $\psi $ in Expansion (\ref{LEMMA2:EXP1}) is the potential due to a set of masses $\{q_1, q_2,...,q_n\}$,
	the truncation error bound of Expansion (\ref{LEMMA2:EXP2}) will share the same form with the one in Equation (\ref{MUL_ERR}) as a result of the uniqueness of the multipole expansion \citep[see][]{greengard1987fast}. The expansions in the following lemmas have similar truncation error bounds, so we will not mention them. 
	\subsection{The FMM-act and Down-pass}
	Another strategy is to convert the multipole expansion of a source box $\mathcal B_S$ into a \textit{local expansion} in the center of any target box $\mathcal B_T$ far from $\mathcal B_S$, which is referred as \textit{$\mathcal B_S$ FMM-acts on $\mathcal B_T$}. Unlike the multipole expansion, the local expansion is valid for the neighborhood of the expansion center with the basis $z^l(\theta,\phi)$ rather than $1/z^l(\theta,\phi)$. One can understand it as an extension of Tyler's series.
	Similar to up-pass, we can pass the local expansion of a box down to its child boxes by translating the center of local expansion to the centers of the child boxes. This processing is called \textit{down-pass}.
	
	Two lemmas similar to Lemma \ref{LEMMA2} can be found to realize FMM-action and down-pass, respectively:
	\begin{lemma}\label{LEMMA3}
		(Conversion of a multipole expansion into a local expansion) $K_S$, $K_T$, 
		$ (\theta^{\sss (S)}_{\sss T}, \phi^{\sss (S)}_{\sss T})$, 
		$ (\theta^{\sss (T)}_{\sss S}, \phi^{\sss (T)}_{\sss S})$, 
		$ (\theta^{\sss (S)},\phi^{\sss (S)})$, 
		and $ (\theta^{\sss (T)}, \phi^{\sss (T)})$ defined as them in Lemma \ref{LEMMA2}, then the Expansion  (\ref{LEMMA2:EXP1}) that converges for $\theta > \theta^{\sss (S)}_r$ in $K_S$ can be expanded as a local expansion in $K_T$
		\begin{align}
			\psi' (\theta^{\sss (T)},\phi^{\sss (T)}) & = C' + Q' \log (1 + \cos\theta^{\sss (T)}) \nonumber \\
			& + 2 \Re \sum_{l=1}^\infty A'_l z^l (\theta^{\sss (T)},\phi^{\sss (T)}) \label{LEMMA_EXP3}
		\end{align}
		that converges for $0< \theta^{\sss (T)} < \theta_r^{\sss (T)} \equiv  \theta_{\sss T}^{\sss (S)} - \theta^{\sss (S)}_r$, where
		\begin{align}
			&C' = C + 2 Q \log\sin\frac{\theta_{\sss T}^{\sss (S)}}{2} +
			2 \Re \sum_{l=1}^{\infty}e^{-i l \phi^{\sss (S)}_{\sss T} } A_l \cot^l\frac{\theta_{\sss T}^{\sss (S)}}{2}\ , \nonumber \\
			&Q' = Q\ ,\ \nonumber  \text{and}\\
			&A'_l = \left ( - \frac{Q}{l}\cot^l \frac{\theta_{\sss T}^{\sss (S)}}{2} + 
			\sum_{l'=1}^\infty e^{-i l' \phi^{\sss (S)}_{\sss T}} A_{l'} T_{ll'} \right) e^{-i l\phi^{\sss (T)}_S} \label{LEMMA_EXP3_2}
		\end{align}
		with
		\begin{align}
			&T_{l l'} = \sum_{n=1}^{\min (l, l') } C_{l-1}^{n-1} C_{l'}^n \cot^{l+l' - 2n} \frac{\theta_{\sss T}^{\sss (S)}}{2} \left (1 + \cot^2\frac{\theta_{\sss T}^{\sss (S)}}{2}\right)^n\ .
		\end{align}
	\end{lemma}
	
	Different from the plane case, in which the down-pass can be achieved by the complete Horner's scheme, on the sphere,  the translation of a local expansion will be more complex. Similar to Lemma \ref{LEMMA2} and Lemma \ref{LEMMA3}, we have

	\begin{lemma} \label{LEMMA4}
		(Translation of a local expansion) $K_S$, $K_T$, 
		$(\theta_{\sss T}^{\sss (S)}, \phi_{\sss T}^{\sss (S)})$,
		$ (\theta_{\sss S}^{\sss (T)},\phi_{\sss S}^{\sss (T)})$, 
		$ (\theta^{\sss (T)},\phi^{\sss (T)})$, 
		and $ (\theta^{\sss (S)}, \phi^{\sss (S)})$
		defined as them in Lemma \ref{LEMMA2}, then the local expansion in $K_s$
		\begin{align}
			\psi (\theta^{\sss (S)},\phi^{\sss (S)}) & = C + Q\log (1 + \cos\theta^{\sss (S)}) \nonumber \\
			& + 2 \Re \sum_{l=1}^\infty A_l z^l (\theta^{\sss (S)},\phi^{\sss (S)})
		\end{align}
		that converges for $0 < \theta^{\sss (S)} < \theta^{\sss (S)}_r$ can be expanded  in $K_T$ as
		\begin{align}
			\psi' (\theta^{\sss (T)},\phi^{\sss (T)})  & = C' + Q' \log (1 + \cos\theta^{\sss (T)}) \nonumber \\
			& + 2 \Re \sum_{l=1}^{\infty} A'_l z^l (\theta^{\sss (T)},\phi^{\sss (T)})\ ,\label{LEMMA_EXP4}
		\end{align}
		which converges for $ 0<\theta^{\sss (T)}<  \theta_r^{\sss (T)} \equiv \theta^{\sss (S)}_r - \theta_{\sss T}^{\sss (S)}$, where
		\begin{align}
			&C' = C + 2 Q \log\cos\frac{\theta_{\sss T}^{\sss (S)}}{2} + 
			2\Re \sum_{l=1}^\infty e^{i l \phi^{\sss (S)}_{\sss T}} A_l \tan^l\frac{\theta_{\sss T}^{\sss (S)}}{2}\nonumber \ ,\\
			&Q' = Q\nonumber ,\ and\\
			&A'_l =  (-1)^l \left ( - \frac{Q}{l}\tan^l\frac{\theta_{\sss T}^{\sss (S)}}{2}+
			\sum_{l'=1}^\infty e^{i l' \phi^{\sss (S)}_{\sss T}} A_{l'} T_{ll'}  \right) e^{-i l \phi^{\sss (T)}_S}
		\end{align}
		with $T_{ll'}$ defined as it in Lemma \ref{LEMMA2}.
	\end{lemma}
	
	With the two above lemmas, we begin to calculate the local expansions from top to down. For the root, we set the local expansion to zero. Then for each non-root level, the local expansion of any box $\mathcal B$ should consist of two parts: One part inherits from  $\mathcal B$'s parent by with down-pass procedure; Another part is converted from the multipole expansions of the  $\mathcal B$'s so-called satellite boxes by FMM-action, where satellite boxes are these far from $\mathcal B$, but not far from $\mathcal B$'s parent.  
	Obviously, the summation of the two parts gives the potential generated by $\mathcal B$'s all far boxes. 
	Finally, we will reach the leaf level boxes, say $\mathcal B_L$. The local expansion of $\mathcal B_L$  gives the contribution from all the far boxes of it. Note that in addition to the local expansion, we also need to obtain the information of $\mathcal B_L$'s near leaf boxes to do the direct summation. In the implement, we build a close-boxes list for $\mathcal B_L$ and when we add a close box $\mathcal B'_L$ into the list, we say \textit{$\mathcal B'_L$ direct-acts on  $\mathcal B_L$}.
	\subsection{Tree-walk Algorithm }
	\newcommand{\forcond}{$i=0$ \KwTo $n$}
	\SetKwProg{Fn}{Function}{}{end}\SetKwFunction{FRecurs}{INTERACTION}
	
	\begin{algorithm}[H]
		\caption{Tree-walk algorithm used by the sFMM}  \label{TREE_WALK}
		\Fn(){\FRecurs{$\mathcal B_S$, $ \mathcal B_T$}}{
			\uIf{$\mathcal B_S$ == $\mathcal B_T$}
			{
				\If{$\mathcal B_S$ is not leaf node}
				{
					\For{ each child i of $ \mathcal B_S $ }
					{
						\For{ each child $j$ of $\mathcal B_S$}
						{
							\texttt{INTERACTION($i$, $j$)}
							
						}
					}
				}
			}
			\uElseIf{$\mathcal B_T$ is far from $\mathcal B_S$}
			{
				$B_S$ FMM-acts on $B_T$
			}
			\uElseIf{neither $\mathcal B_S$ nor $\mathcal B_T$ is the leaf}
			{
				\tcc{$r(\mathcal B)$ gives the radius of $\mathcal B$}
				\If{ $r(\mathcal B_S) > r(\mathcal B_T)$ }
				{
					\For{each child $i$ of $\mathcal B_S$}
					{
						\texttt{INTERACTION($i$, $\mathcal B_T$)}
					}
				}
				\Else
				{
					\For{each child $j$ of $\mathcal B_T$}
					{
						\texttt{INTERACTION($\mathcal B_S$, $j$)}
					}
				}
			}
			\uElseIf{$\mathcal B_S$ is not the leaf}
			{
				\tcc{$\mathcal B_T$ is the leaf}
				\For{each child $i$ of $\mathcal B_S$}
				{
					\texttt{INTERACTION($i$, $\mathcal B_T$)}
				}
			}
			\uElseIf{$\mathcal B_T$ is not the leaf}
			{
				\tcc{$\mathcal B_S$ is the leaf}
				\For{each child j of $\mathcal B_T$}
				{
					\texttt{INTERACTION($\mathcal B_S$,$j$)}
				}
			}
			\Else(\tcc{both $\mathcal B_S$ and $\mathcal B_T$ are the leaves})
			{
				$\mathcal B_S$ direct-acts on $\mathcal B_T$
			}
			\Return 
		}
	\end{algorithm}
	
	As described above, the actions (including FMM-action and direct-action) and pass-down can be completed together. 
	However, in practice, before doing the down-pass, all the actions processes can be done by using a tree-walk algorithm modified from \cite{dehnen2002hierarchical}, which is coded as a function \texttt{INTERACT} in Algorithm \ref{TREE_WALK}.

	Note that the function \texttt {INTERACTION} calculates the action of source box $\mathcal B_S$ on target box $\mathcal B_T$ recursively. If $\mathcal B_S$ and $\mathcal B_T$ is the same box, it will calculate the interaction inner $\mathcal B_S$ (or $\mathcal B_T$). 
	One can finish all the action works by calling \texttt {INTERACTION}($\mathcal B_R$, $\mathcal B_R$), where $\mathcal B_R$ is the root of the tree.
	\subsection{The Multipole Acceptance Criterion (MAC)}\label{SFMM:MAC}
	Only when the target box is \textit{far} from the source box, Expansion  (\ref{LEMMA_EXP3}) can converge fast. 
	The Multipole Acceptance Criterion (MAC) \cite[see, e.g.,][]{dehnen2002hierarchical} provides a mathematical definition of \textit{far}.

	We define
	the radius of a box as the angular radius of the smallest circle on the sphere that can cover the box, and
	the distance between two boxes as the angular distance of their centers. The first MAC is 
	\begin{align}
		d - R_T - c_s R_S > 0\ ,\label{MAC1}
	\end{align}
	where  $c_s > 1 $ is a constant, $R_T$, $R_S$, and  $d$ are the radius of target box, the radius of source box, and the distance between two boxes, respectively. The $c_s$ sets a protected circular area with radius of $c_sR_S$ around the source box.  If the target box intersects with this protected area, it will be classified as the close box of the source box. The reason to set such a protected area is that, sometimes, the particle is viewed as a mass distribution.
	So there may be some mass contributed from a particle, even out of the box it belongs to.
	In our code, we set $c_s=2$.
	
	However, the first MAC is not enough, because one will encounter a case that $R_T \approx d$, where the first MAC is fitted, but Lemma \ref{LEMMA3} still doesn't converge fast on the border of the target box. So we need a second MAC
	\begin{align}
		R_T < c_t d\ ,\label{MAC2}
	\end{align}
	where $c_t < 1$ is a constant. A smaller choice of $c_t$ guarantees faster convergence even on the border of the target box, however, it increases the times to do FMM-action. We set $c_t = 0.5$ in practice.
	
	If Equations (\ref{MAC1}) and  (\ref{MAC2}) hold, we will say that the target box is \textit{far} from the source box.
	\subsection{Getting Lensing Quantities }\label{FMM:GET_LQ}
	After the sFMM procedure (i.e., up-pass, actions, and down-pass),
	the lensing potential in each leaf box can be written in the local coordinate system $K_L$ as
	\begin{align}
		\psi & =\psi_i + \psi_{e}  \nonumber \\
		&\approx \psi_i +  C + Q\log (1 + \cos\theta^{\sss (L)}) \nonumber \\
		&+ 2 \Re \sum_{l=1}^p  A_l z^l (\theta^{\sss (L)},\phi^{\sss (L)}) \ ,
	\end{align} where $(\theta^{\sss (L)}, \phi^{\sss (L)})$ is the coordinate of a target point in $K_L$, $\psi_i$ and $\psi_e$ are the potentials coming from the near particles through direct summation and the far particles through the local expansion, respectively.
	
	From Equation  (\ref{ALP_PSI}) and  (\ref{GAM_DEF}) one can calculate $\bm \alpha$ and $\bm \gamma$ conveniently in the local spherical coordinate systems, for example, the $\bm e_\theta^{\sss (L)}$ component of $\bm\alpha$ is  
	\begin{align}
		\alpha_\theta^{\sss (L)} &\approx  \partial_\theta^{\sss (L)} \psi_i + \partial_\theta^{\sss (L)}\psi_e \nonumber \\
		&  = \partial_\theta^{\sss (L)}\psi_i - Q \tan\frac{\theta^{\sss (L)}}{2} \nonumber  + 2\Re \sum_{l=1}^p A_l\partial_\theta^{\sss (L)} z^l (\theta^{\sss (L)},\phi^{\sss (L)})\nonumber  \\
		&  = \partial_\theta^{\sss (L)}\psi_i - Q \tan\frac{\theta^{\sss (L)}}{2}  + 2\Re \sum_{l=1}^p \frac{l}{\sin\theta^{\sss (L)}}A_l z^l (\theta^{\sss (L)},\phi^{\sss (L)}) \ ,
	\end{align}where $\psi_i$ can be obtained from NPM analytically. 
	
	Note that the components of $\bm \alpha $ and $\bm \gamma$ in the global coordinate system can be obtained through coordinate transformations.
	\subsection{Overview and Time Complexity Analysis}
	Now we give an overview of the algorithm. The sFMM is roughly divided into five parts.
	Firstly, we divide a sphere into a quasi-quad-tree according to the particle distribution on the sphere. 
	Secondly, we build the multipole expansions for all the leaf boxes with Lemma \ref{LEMMA1} and pass up these expansions from down to top with Lemma \ref{LEMMA2}. After that, every box in all levels gets a multipole expansion.
	Thirdly, we do FMM-action and direct-action by using Algorithm \ref{TREE_WALK} and Lemma \ref{LEMMA3}. Then every non-root box gets a local expansion from its satellite boxes. In addition, each leaf box gets a list of its near boxes. 
	Fourthly, we do down-pass from top to down, until meet with the leaf boxes. 
	After finishing this, the local expansion of each leaf box gives the potential generated by its all-far boxes. 
	Finally, for each target point, we find the leaf box it belongs to, then calculate the lensing quantities by combining the far particles' local expansion and the near particles' direct summation.
	
	To analyze the time cost of the sFMM, a lot of factors have to be taken into account. However, some factors are actually hard to evaluate accurately.  
	For instance, because of the adaptive division technique we use, the tree data structure will rely on the particle distribution, which increases the complexity to analyze. 
	Meanwhile, among these factors, our main concern is the relation between the time cost and particle number $N$.
	So, to simplify our analysis, we set particles evenly distribute on the sphere and $m=1$.
	
	Under these assumptions, firstly, one can find that  the time cost of dividing the sphere and assigning particles is 
	\begin{align}
		T_0 \approx c_0 N\log_4 N \label{T0}\ ,
	\end{align}
	where the constant $c_0$ is the time spent to assign a particle to the child box and $\log_4 N$ is the depth of the tree.
	Secondly, the time to do the sFMM procedure is given by
	\begin{align}
		T_1 \approx c_1 N + \frac{4}{3}c_1' N + \frac{4s}{3} c_1'' N + \frac{4}{3} c_1''' N\ ,
	\end{align}
	where $s$ is the average number of a box's satellites, $c_1$,$c_1'$,$c_1''$ and $c_1'''$  are the times cost to build the multipole expansion for one particle, to pass up one multipole expansion, to expand one multipole expansion to local expansion, and to pass down one local expansion, respectively. As expected, the total time cost of this part is $O (N)$. A more complete analysis of this part was given by \cite{carrier1988fast}. 
	Lastly, the time cost to get results from leaf boxes will be
	\begin{align}
		T_2 \approx  c_2  M \log_4 N+ c_2' M,\label{T2}
	\end{align}
	where $M$ is the number of target points, $c_2 \log_4 N $ is the time spent to find the leaf box that a target particle belongs to, and $c_2'$ is the time cost to get the lensing quantities at one target point.
	
	In practice, $T_0$ is much smaller than $T_1$. For example,  $T_0/T_1 \approx 0.015$  with $N=10^7$ in our test.
	Besides, $c_2'\gg c_2$ in Equation (\ref{T2}), so for realistic use, the total time cost is about
	\begin{align}
		T_{sFMM} \approx c_N N + c_M M\ ,
	\end{align}
	where $c_N = c_1 + (c_1'+c_1'') + 2c_1'''$  and $c_M=c_2'$. So the sFMM is an almost linear method.
	In Section \ref{TEST} we will verify it.

	
	\section{Result}\label{TEST}
	In this section, we carefully test the performance of the sFMM. Focusing on the algorithm itself, we set the number of layers $N_L=1$ and take the space-curvature $K=0$. Note that such a setting will not influence the performance of the algorithm.
	\subsection{Precision and Time Cost of the sFMM}
	\label{TEST_S1}
	\begin{figure*}
		\centering
		\subfigure
		{
			\includegraphics[width=\linewidth]{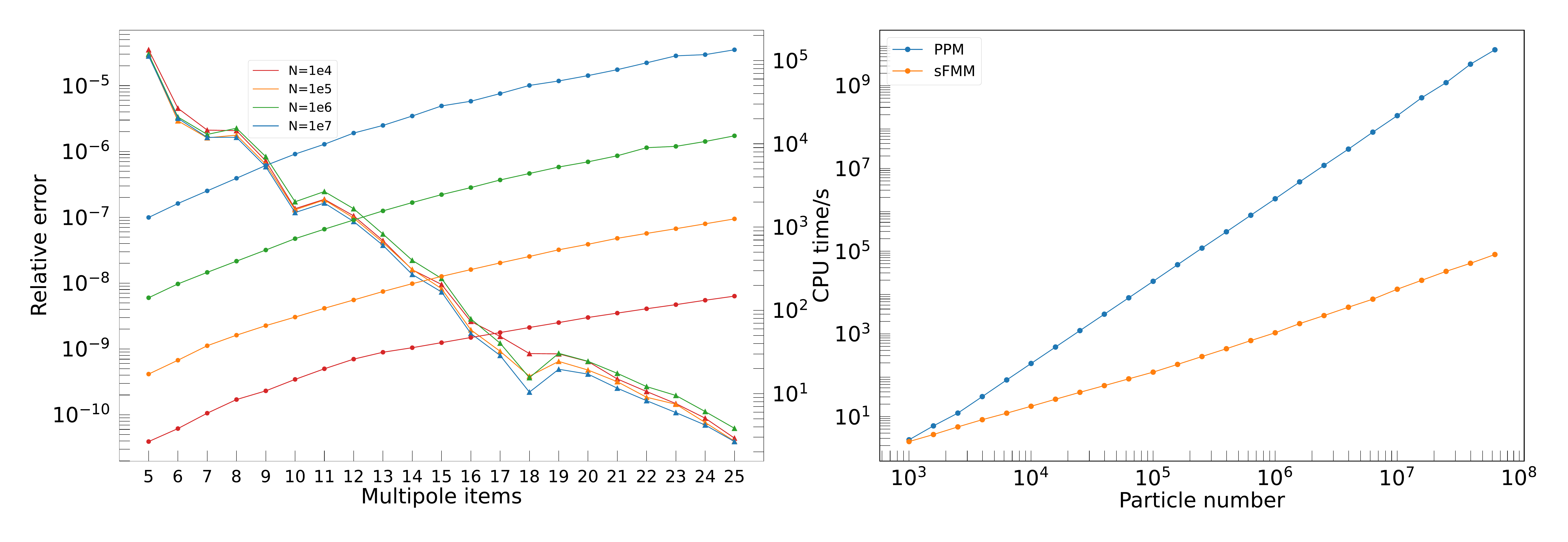}
		}
		\caption{Left: the relative errors (lines with triangles) and time costs (lines with dots) of the sFMM. 
			In the test, $m$, the maximal number of particles in leaf boxes is fixed to 1; The MAC parameters $c_s$ and $c_t$ are fixed to 2.0 and 0.5, respectively.
			Right: The time costs of the sFMM and PPM. 
			In the test of the sFMM, the truncation parameter $p$ is fixed to 10 and other parameters are set as them in Left. Such a setting leads to the precision around $10^{-7}$.}
		\label{TEST_T}
	\end{figure*}
	To test the precision of the sFMM, we first randomly scatter $N$ sources with random mass on a sphere, then construct the tree and do the FMM procedure.
	Next, $M$ target points on the sphere are randomly selected to obtain the relative error of lensing potential by $|(\psi^{\sss (S)} - \psi^{\sss (P)})/\psi^{\sss (P)}|$, where $\psi^{\sss (S)}$ and $\psi^{\sss (P)}$ are respectively calculated from the sFMM and the particle-point method (PPM). 
	Here PPM sums the contributions from all sources directly, thus it has the machine precision and time complexity of $O (NM)$. 
	Considering the time cost of PPM, we only assess the relative error for a set of test points with $M=M_0=1000$.
	
	In practice, it is more meaningful to consider the time cost in the situation of $M=N$. Since we just use $M_0$ test points, it can be estimated by
	\begin{align}
		T_{sFMM} =  T_0 + T_1 + \frac{T_{\sss M_0}}{M_0}N
	\end{align}where $T_{\sss M_0} $ is the total time spent to get the results on these $M_0$  target point from leaf boxes (described in Section \ref{FMM:GET_LQ}), $T_0$ and $T_1$ are the time to construct the tree and do FMM procedure, respectively.
	Similarly, the time cost of PPM is 
	\begin{align}
		T_{PPM} = \frac {T^{\sss (P)}_{\sss M_0}}{M_0} N \ ,
	\end{align}where $T^{\sss (P)}_{\sss M_0}$ is the time spent to get the results on these $M_0$ test points with PPM.

	Changing $N$ from $10^4$ to $10^7$ and truncation parameter $p$ from 5 to 25,
	we get the left panel of Figure \ref{TEST_T}.
	The result indicates that the sFMM can achieve high calculation accuracy and good numeric stability. For different numbers of point sources, the relative errors can be steadily suppressed as $c^{p}$.
	The fitting of the precision-$p$ curves gives $c\approx 0.636\pm 0.026$, where $0.026$ is the standard deviation error.
	Specifically, the error is under $10^{-4}$ for $p=5$ and  reaches $10^{-10}$ for $p=10$.
	
	At the same time, one should note that the time spent in the sFMM increases as $p^n$ and theoretically, $n=3$. The fitting of the time-$p$ curves gives an average value of $n\approx 3.013 \pm 0.065$, as expected. Taking $N=10^7$ as the example, the time increases from  about $10^3$ seconds for $p=5$  to  about $10^5$ for $p=25$. In practice, one should trade off the precision and time, then choose an appropriate truncation parameter.
	
	In the right panel of Figure \ref{TEST_T}, we compare the time costs of the sFMM and PPM for $N$ from $10^3$ to about $6\times10^7$. With $N=10^3$, the time costs of both the sFMM and PPM are about $3$ seconds. But with increasing $N$, PPM's time cost increases much faster than the sFMM's. The fitting of the two curves gives the time dependence of  $N^{0.937\pm 0.010} \approx N$ for the sFMM and $N^{1.989\pm 0.005}\approx N^2$ for PPM, as expected.  When $N$ is around $6\times 10^7$, the PPM's time cost is $T_{PPM} \approx 7.4\times 10^9$ seconds while the sFMM's is $T_{sFMM} \approx 8.3\times 10^4$ seconds, leading to a ratio ${T_{PPM}}/{T_{sFMM}} \approx 8.89\times 10^4$.

	\subsection{Comparing with the FSHT}

	\begin{figure*} 
		\centering 
		\subfigure
		{
			\includegraphics[width=.9\linewidth]{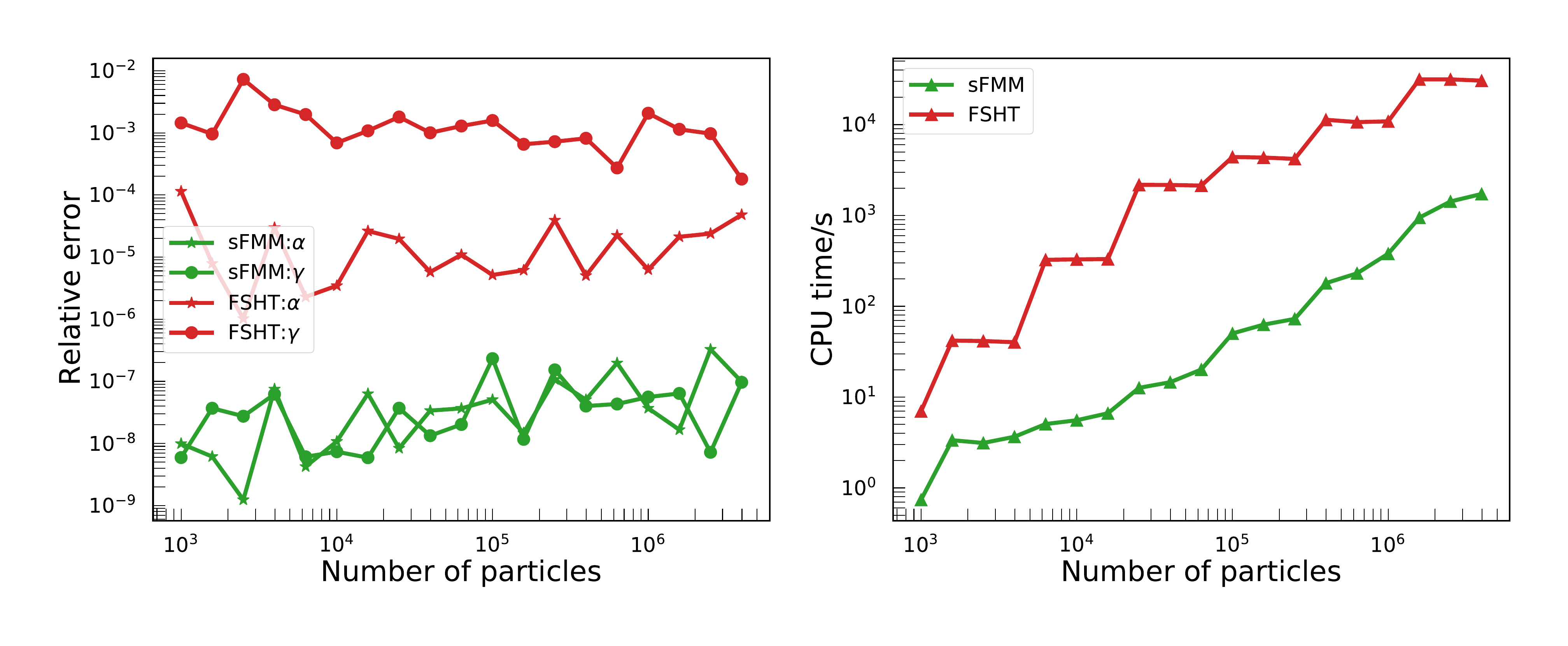}
		}
		\caption{Left:  The precision of the sFMM and the FSHT. Right: The time cost of the sFMM and the FSHT. }
		\label{NTEST}
	\end{figure*}
	\begin{figure*}
		\centering
		\subfigure[$\gamma$ from the sFMM for $\sigma = 0.1$]
		{
			\includegraphics[width=.3\linewidth]{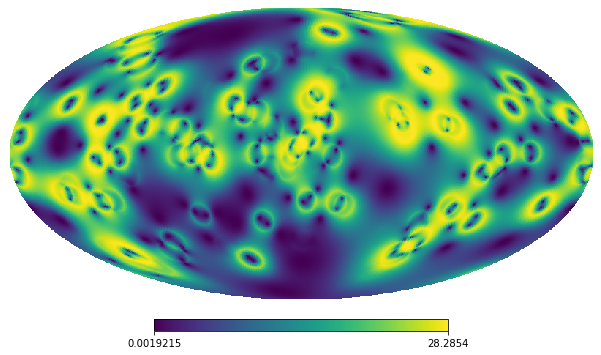}
			\label{compare:gamma_s1}
		}
		\quad
		\subfigure[$\gamma$ from the sFMM for $\sigma = 0.03$]
		{
			\includegraphics[width=.3\linewidth]{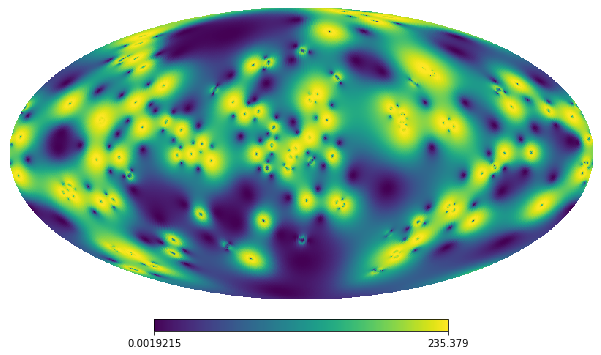}
			\label{compare:gamma_s2}
		}
		\quad 
		\subfigure[ $\gamma$ from the sFMM for $\sigma = 0.001$]
		{
			\includegraphics[width=.3\linewidth]{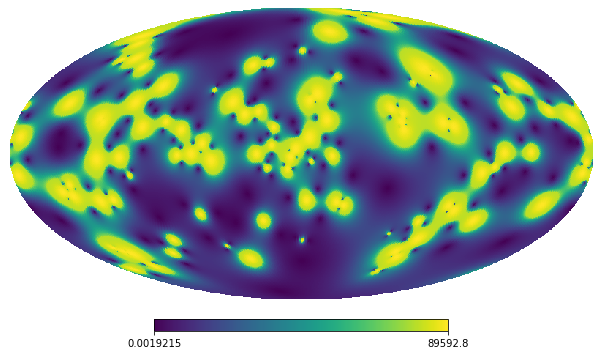}
			\label{compare:gamma_s3}
		}\\
		\subfigure[ $\gamma$ from the FSHT for $\sigma = 0.1 $]
		{
			\includegraphics[width=.3\linewidth]{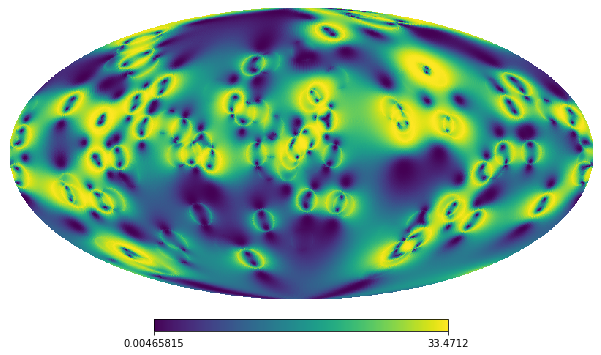}
			\label{compare:gamma_f1}
		}
		\quad
		\subfigure[$\gamma$ from the FSHT for $\sigma = 0.03$]
		{
			\includegraphics[width=.3\linewidth]{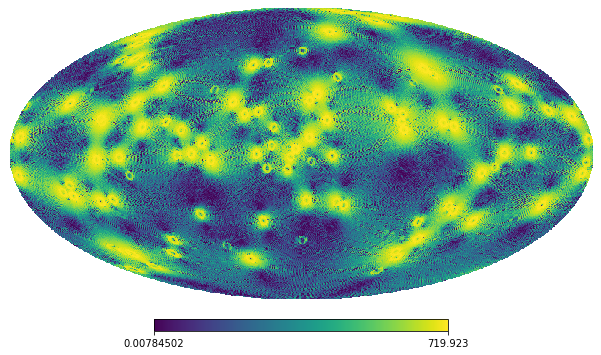}
			\label{compare:gamma_f2}
		}
		\quad
		\subfigure[$\gamma $ from the FSHT for $\sigma = 0.001$]
		{
			\includegraphics[width=.3\linewidth]{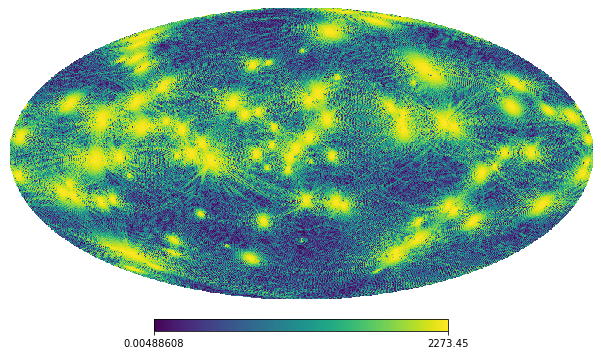}
			\label{compare:gamma_f3}
		}
		\label{justsmooth}
		\caption{ The module of shear $\gamma \equiv  \sqrt{\gamma_1^2 + \gamma_2^2}$ from the sFMM and the FSHT for different particle radii $\sigma$s. In the test, the sFMM parameters are fixed to $p = 10$, $m=1$, $c_s=2$, and $c_t=0.5$; the FSHT parameters are fixed to $N_{pix}=196608$ and $l_{max}  = 3N_{side}-1 = 383$. Note that in (e), $\sigma = 0.03 \approx \pi/N_{side}$.
		}
		\label{COMPARE}
	\end{figure*}
	To compare with the FSHT, one needs to interpret a particle as a point-like mass that occupies a small space with a certain profile (i.e., NPM in Section \ref{BasicIdea}), for instance, a Gaussian or triangular one. 
	In the tests on both the sFMM and the FSHT, a generalized Epanechnikov kernel
	\begin{align}
		W (\theta) = 
		\left\{
		\begin{matrix}
			\frac{1}{\mathcal N_\sigma} \left[ 1 - \left (\frac{\sin(\theta/2)}{\sin(\sigma/2)}\right)^2 \right] & \quad  \theta \leq \sigma \\
			0 &  \quad \theta>\sigma
		\end{matrix} \label{KERNEL}
		\right. 
	\end{align}is employed as the particle profile, because it is compact and computationally efficient to implement,
	where  $\sigma$ is the length of particles and $\mathcal N_\sigma = 2\pi \sin^2(\sigma/2)$ is the normalization factor. We set $\sigma = \sqrt{4\pi/N}$ for $N$ particles. 
	
	In the test using the sFMM, the truncation parameter $p$, the minimum number of particles in leaf box $m$, and the MAC parameters $c_s$ and $c_t$ are set to 10, 1, 2, and 0.5, respectively.    
	The lensing quantities of a smoothed particle with profile (\ref{KERNEL}) can be obtained analytically.    For example, the deflection angle for Epanechnikov kernel (\ref{KERNEL})  is
	\begin{align}
		\bm \alpha(\theta) =
		\left\{
		\begin{matrix}
			\frac{q}{8 \sin^4(\sigma/2)} \sin(\theta) \left[ 1 - 2 \cos(\sigma) + \cos(\theta) \right]  \hat{\bm\alpha}& \quad  \theta \leq \sigma \\
			\frac{q}{2}\cot(\theta/2) \hat{\bm\alpha}&  \quad \theta>\sigma \label{KERNEL_ALPHA}
		\end{matrix}
		\right.    \ ,
	\end{align}
	where $q$ is the mass of the particle and $\hat{\bm\alpha}$ is the unit vector on the sphere, which points to the particle from the target point.    With Equation (\ref{KERNEL_ALPHA}), one can directly calculate the deflection angles contributed by near-field particles and sum them according to the superposition principle.

	In the test with the FSHT, each particle is assigned onto the mesh by
	\begin{align}
		\rho_{ij} =\frac{q_i}{ \Omega_{grid} \sum_{k=0}^{N_{grid}-1} W (|\bm\theta^{ (p)}_i - \bm\theta^{ (g)}_k|)} W (|\bm\theta^{ (p)}_i - \bm\theta^{ (g)}_j|)\ ,
	\end{align}
	where $\rho_{ij}$ is the contribution of the $i$-th particle on the $j$-th grid, $q_i$ is the mass of the $i$-th particle, $N_{grid}$ is the number of grid, $\Omega_{grid}$ is the area of each grid pix, $\bm\theta^{ (p)}_i$ is the position of the $i$-th particle, and $\bm\theta^{ (g)}_j$ is the position of the $j$-th grid. The normalization factor under $q_i$ is used to assure the conservation of the total mass.
	To improve the precision of the FSHT, we set the grid length $d_{grid} \approx \sigma/16$, which is much smaller than $\sigma$.
	We also set the degree of spherical harmonics up to $ l_{max} = \min (2700,3N_{side})$,  where 2700 is the max degree that the associated Legendre polynomials can be stably computed in GNU Scientific Library\footnote{\url{https://www.gnu.org/software/gsl}}, $N_{side} = \sqrt{N_{grid}/12}$ is the HEALPix parameter that defines the number of divisions along each side of a base box that is needed to reach a desired high-resolution partition (corresponding to the Nyquist frequency $\omega_{\sss N} \approx 2N_{side}$).
	
	We randomly scatter from $10^3$ to $10^6$ particles with random masses on the sphere, solve the PE, and obtain the errors relative to PPM and the time costs of the two methods. Note that the time cost of the FSHT includes the time spent to assign particles onto the mesh. The results are shown in  Figure \ref{NTEST}. 
	In the sFMM, the precision of $\alpha$ and $\gamma$ field is stable as about $10^{-7}$, as expected. 
	For the FSHT, the precision of $\alpha$ and $\gamma$ field is about $10^{-5}$ and $10^{-3}$, respectively.  
	In general, the sFMM's precision is better than the FSHT's in our test.
	Meanwhile, the sFMM's time cost is cheaper than the FSHT (see the right panel of Figure \ref{NTEST}). On average, the time spent by the FSHT is about 100 times of that by the sFMM. Note that the trapezoidal platforms in the right panel are formed because we take a smaller step of the particle numbers compared with the test in Section \ref{TEST_S1}. Almost the same number of particles will lead to the same time cost since trees with the same depth will be constructed in the sFMM and the same $N_{grid}$ will be taken in the FSHT. 
	
	The two methods' ability to obtain high-frequency information should be compared. To gain some instinct on the comparison between the FSHT and the sFMM, we fix other parameters and solve the PE with 100 particles on a sphere with different particle radii. 
	The $\gamma$ fields in Figure \ref{COMPARE} illustrate the limitation of the FSHT to obtain high frequency components:
	As the radius of these particles decreases, the sFMM can always give an accurate result, while the FSHT gives one with visible noise. The result is reasonable since the smaller the particles are, the higher the proportion of high-frequency components, leading to the increase of the FSHT's truncation error.

	Here we compare the sFMM and the FSHT from a theoretical point of view. Both of them expand the field with some base functions: the sFMM takes $z^l (\theta,\phi) = \tan^l\left ({\theta}/{2}\right)e^{il\phi}$ as base functions, while the FSHT takes the spherical harmonics $Y_l^m (\theta,\phi)$. But the sFMM expands the field with $z^l(\theta,\phi)$ locally; the FSHT does it with $Y_l^m (\theta,\phi)$ globally. Besides, from an informational point of view, the more expanding items we use, the more information about the field we get. The number of expanding items in sFMM is about $p$ times the number of leaf boxes, which is much more than it in the FSHT (up to 2700 in our test), and that's why we can get a more accurate result from the sFMM.
	So for a globally \textit{low-frequency} mass distribution, instead of a locally point-like mass one in the test, the FSHT will behave much better. For example, we can solve the PE of mass distribution $\kappa(\bm\theta) = Y_1^0(\theta,\phi) = \sqrt{{3}/{(4\pi)}}\cos\theta$ with the FSHT, then the precision will be about $10^{-7}$ for $N_{side}= 2^7$. For the sFMM, to solve the PE with such a mass distribution, one needs to sample it with particles, which will introduce noise.
	We will test this noise in Section \ref{SEC_NFW}.
	\subsection{Test with a Navarro–Frenk–White  (NFW) Profile}\label{SEC_NFW}
	\begin{figure*}[ht]
		\centering
		\subfigure
		{
			\includegraphics[width=.75\linewidth]{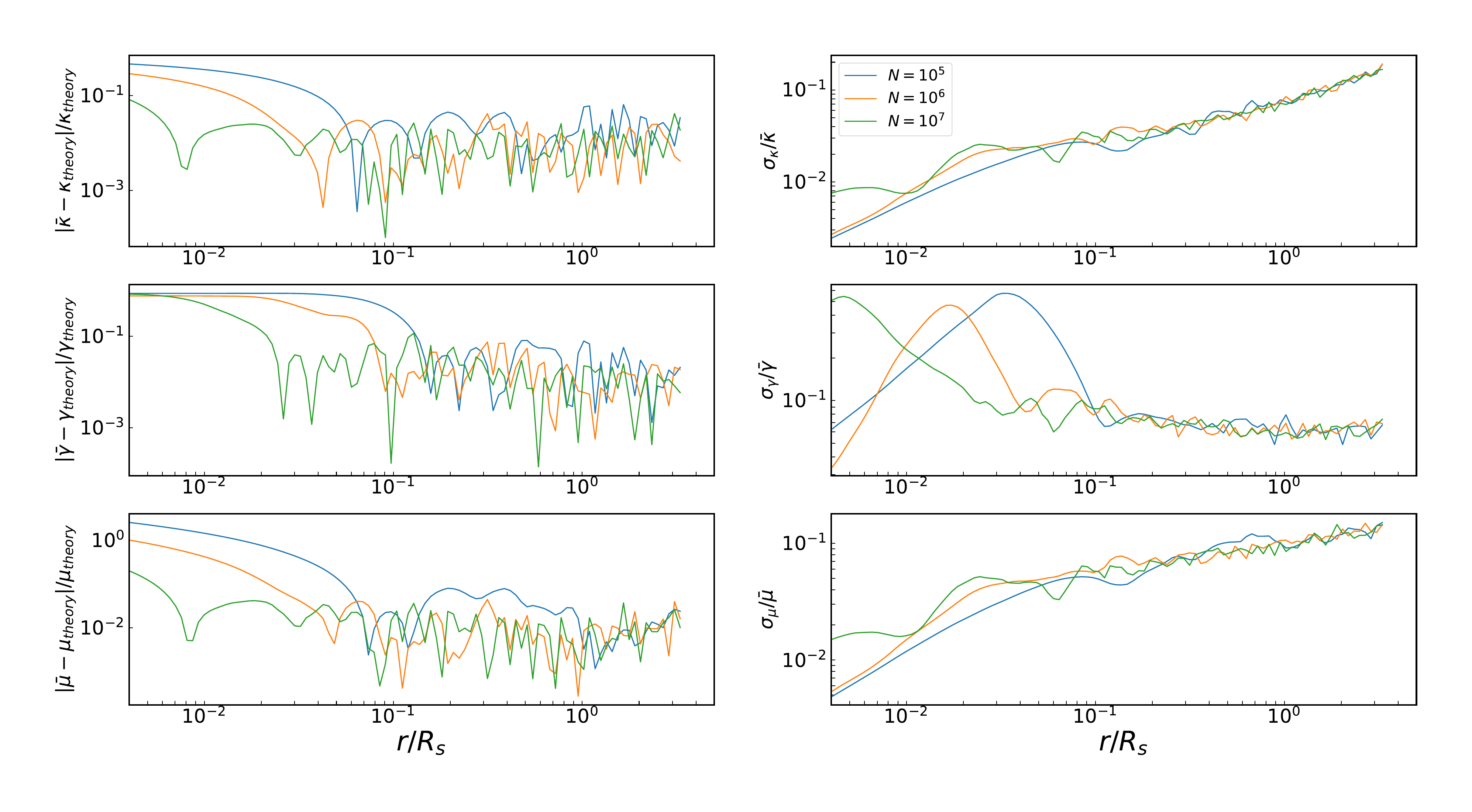}
		}
		\caption{ The relative error and standard deviation of $\kappa$ (top), $\gamma$ (middle), and $\mu$ (bottom). The standard deviation is normalized by the mean value. In the test, $\beta$ is set to 5.}
		\label{NFW_N}
	\end{figure*} 
	\begin{figure*}[ht]
		\centering
		\subfigure
		{
			\includegraphics[width=.75\linewidth]{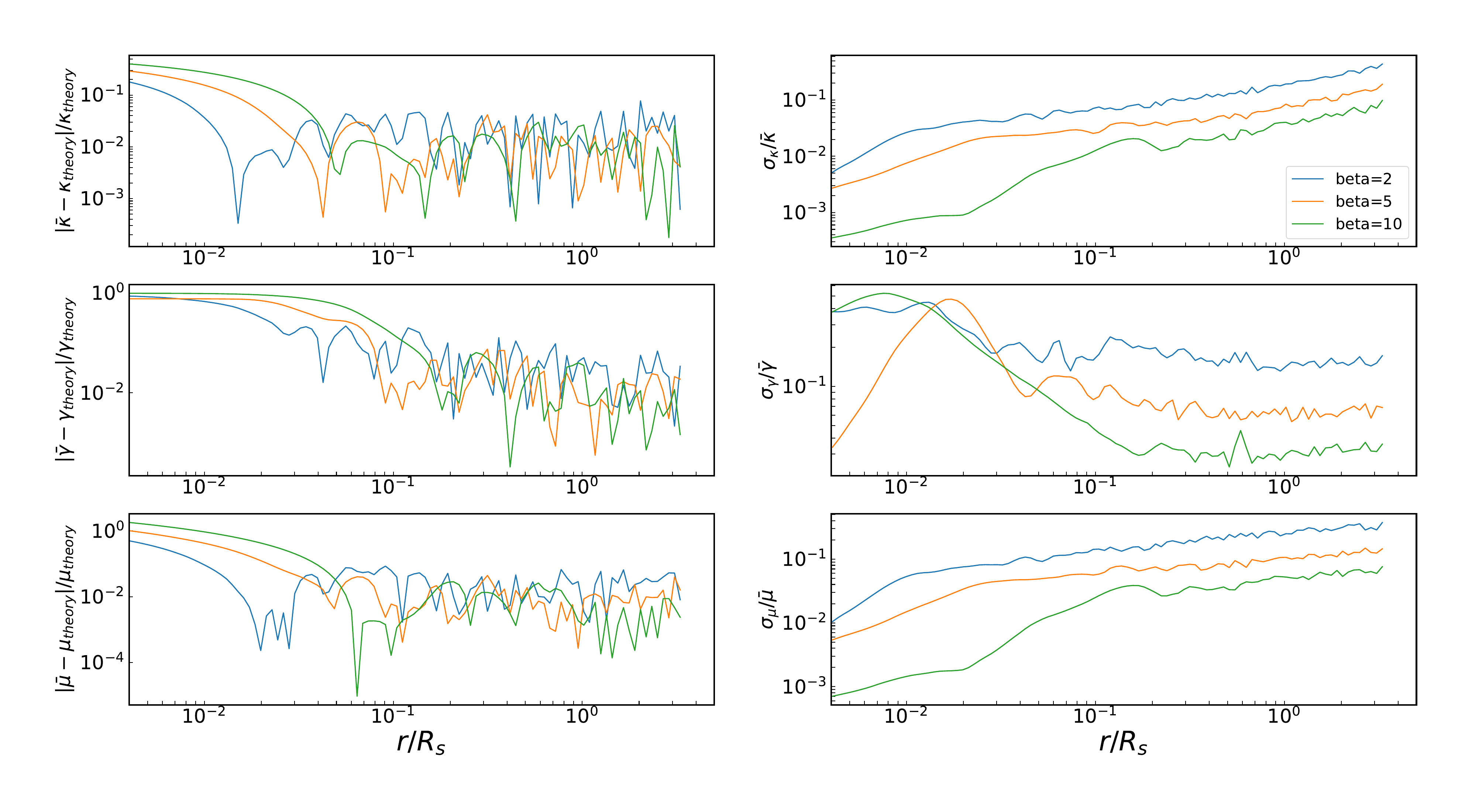}
		}
		\caption{ The same as Figure \ref{NFW_N} but for different $\beta$s. $N$ is set to $10^6$.
		}
		\label{NFW_beta}
	\end{figure*}
	
	The sFMM behaves well in the test with point or point-like source distribution. However, as previously mentioned, to solve a continued mass distribution by the sFMM, one needs to sample it with particles. We quantify the noise from sampling using the NFW \citep{Navarro_1997} profile, which is the case for dark matter distribution of most astronomical objects, such as galaxy and galaxy cluster. The analytic $\kappa$ field of an NFW halo can be found in \cite{wright2000gravitational}
	\begin{align}
		\kappa (x) = \kappa_s \times\left\{
		\begin{matrix}
			&\frac{4}{x^2-1}\left[ 1 - \frac{2}{\sqrt{1-x^2}}\mathrm{arctanh} \sqrt{\frac{1-x}{1+x}}   \right]  & 0<x<1\\
			&\frac{4}{3} & x=1\\
			&\frac{4}{x^2-1}\left[ 1 - \frac{2}{\sqrt{x^2-1}}\mathrm{arctan}\sqrt{\frac{x-1}{x+1}}   \right]&  x>1,
		\end{matrix}
		\right. \label{NFW}
	\end{align}
	where $\kappa_s = \rho_s   R_s /\Sigma_r$ and $x=r/R_s = cr/R_{vir}$ with $\Sigma_r $ the critical over-density of GL, $\rho_s$, $Rs$, $c$, and $R_{vir}$ the scale over-density, scale radius, concentration parameter, and the virial radius of the NFW profile, respectively. Here we set $R_{vir} = 0.709h^{-1}Mpc$, $M_{vir} = \pi \Sigma_r\int_0^{R_{vir}}r\kappa (r/R_s)dr = 10^{14}h^{-1}M_\odot $, $\Sigma_r = 9.93\times 10^{8}hM_\odot Mpc^{-2}$, and use the mass-concentration from \cite{dutton2014cold}.
	In addition, from the $\kappa$ field, we can obtain $\gamma$ by
	\begin{align}
		\gamma (r) = \frac{1}{r^2} \int_0^{r}r'\kappa (r') dr' - \frac{1}{2} \kappa (r)\ ,
	\end{align}
	which is valid for all mass profiles with radial symmetry.
	
	In the test, we use the MCMC method to sample the profile given in Equation (\ref{NFW}).
	Two parameters will influence the noise from sampling. The first one is $N$, the number of sampling particles. 
	Another one is $\sigma$, the radius of NPM, where we use a scale radius $\beta = \sigma /\sqrt{4\pi/N}$ to represent it. 
	The sampling near the center is far from the PDF  because of the $\log (x)$ singularity at $x=0$.  However, such a singularity will hardly influence the lensing potential far from the origin point because  $\lim_{r\to 0} M(r) = 0 $, where $M(r) \equiv \pi \Sigma_r \int_0^r r' \kappa(r')dr'$ is the total mass inside radius $r$.
	

	To test the influence of $N$ on the results, we fix $\beta = 5$ and set $N = 10^5$, $10^6$, and $10^7$, separately, then obtain the average error and mean-value-normalized standard deviation of the $\kappa$, $\gamma$, and $\mu$ field in different radius.  The results were shown in Figure \ref{NFW_N}.
	The test shows that 
	outside $0.1R_s$, $N$ hardly influences the result and the average errors are in the order of $10^{-2}$. 
	However, the standard deviation is not good in the test. For example, the standard deviation of $\kappa$ is about $10^{-1}$ for all $N$s. This is due to the low mass density outside $R_s$ where it can be hardly mimicked by fewer particles. 
	This problem may be relieved by increasing $\beta$.
	To test the influence of $\beta$ on the results, we set $N = 10^6$ and test with $\beta = 2, 5, 10$.
	As shown in Figure \ref{NFW_beta}, larger $\beta$ does decrease the standard deviation. For example, the average standard deviation of $\kappa$ is $0.284$ for $\beta=2$ and $0.057$ for $\beta=10$. 
	
	We note that in this test, the particles are given the same size.   This leds to a relatively weak ability to approximate the real density by sampling.   A better way is to take an adaptive smoothing where the size of a particle is related to the local density, for example the one used in \cite{10.1093/mnras/stu1859} and its implementation GLAMER\footnote{\url{http://glenco.github.io/glamer/}}.   Since this method helps to reduce the particle noise in the low-density region compared with the FFT-based method, one may combine it with the sFMM to get a more precise result.
	\section{Conclusion}
	\label{CONCLUSION}
	In this paper, we have developed an sFMM method, which is an extension of the traditional FMM to sphere $\mathbb S^2$, to solve the spherical Poisson Equation that calls for extensive calculation in ray tracing of gravitational lensing. 
	We have first pointed out the shortcomings of the existing algorithms and the potential advantage of the standard FMM to overcome these shortcomings.
	To extend the standard FMM to a sphere, we derive some formulae in spherical ray-tracing and obtain the Green function of the spherical Laplacian.
	Then, following the idea of the standard FMM, we present four Lemmas and the details of how to implement them into the sFMM.
	In addition, we test the time cost and precision of sFMM and compare them with the FSHT. The results show that the sFMM can achieve better performance in both the calculation speed and accuracy than the FSHT to solve the PE on a sphere.
	In general, we believe that the sFMM is a powerful tool for ray tracing simulation of a large- even full-sky area using N-body simulations. In particular, we hope it can be used to achieve high accuracy for GL simulation involving both weak and strong lensing, even micro lensing. We will investigate its performance in full-sky GL simulation in more detail and release the code in a future paper. 
	
	\begin{acknowledgments}
		This work is supported by the science research grants from the China Manned Space project with NO.CMS-CSST-2021-A03, CMS-CSST-2021-B01, the NSFC (No.11825303, 11861131006, 11903082), the Fundamental research fund for Chinese Central Universities (226-2022-00216), and the cosmology simulation database (CSD) in the National Basic Science Data Center (NBSDC-DB-10). We thank Tingfei Li for helpful discussions.
		
	\end{acknowledgments}
	
	\appendix
	
	\section{Derivation of Lensing Potential}\label{APPENDIX:A}
	In the $i$-th layer, the deflection angle of light is given by
	\begin{align}
		\bm \alpha_i  (\bm\theta_i)
		& =  \int_{r_{i-\frac{1}{2}}}^{r_{i+\frac{1}{2}}}dr \frac{2}{c^2}\bm\nabla_\perp \Phi\left ( \bm\theta(\bm \theta_i, r) ,r\right)\ ,
	\end{align}
	where $\bm\theta$ is the angular position of the photon and $\bm\theta_i$ is the angular position where the photon enters the layer. With Born's approximation, one integrates along the unperturbed path of light inside the layer.
	\begin{align}
		\bm \alpha_i  (\bm\theta_i)
		& \approx  \frac{2}{c^2} \int_{r_{i-\frac{1}{2}}}^{r_{i+\frac{1}{2}}}dr \bm\nabla_\perp  \Phi (\bm\theta_i,r),
	\end{align}
	For a point  mass $M$  at the top of the sphere it gives \citep[see, e.g.,][]{meneghetti2022probability}
	\begin{align}
		\alpha_\theta  &=  \frac{4GM}{c^2b } \cos \frac{\theta}{2} \\
		&= \frac{2GM}{c^2a f_K (r_i)}\cot\frac{\theta}{2} \\
		& = \frac{4GM}{c^2af_K (r_i)} \partial_{\theta}\log \sin\left(\frac{\theta}{2}\right)\ ,
	\end{align}
	where $b$ is the impact parameter, $a(r_i)$ is the scale factor in when the photon is in $r_i$,
	and $\theta$ is the angular distance between the point mass and the photon. Note the factor $\cos\frac{\theta}{2}$ projects the potential gradient to the vertical direction of light.
	For the photon in the layer, we assume that it can only be influenced by the density inside the layer. This assumption will break down for thin layers \cite[as argued in][from different aspect]{das2008large}. Generalize this formula to any mass distribution, we finally get 
	\begin{align}
		\bm \alpha (\bm\theta_i)  =  \bm\nabla_{\bm\theta_i} \int d\Omega' \kappa (\bm\theta') \mathcal G (\bm\theta_i,\bm\theta') = \bm\nabla_{\bm\theta_i} \psi (\bm\theta_i)\ ,
	\end{align}
	where $\kappa$ is the dimensionless mass density defined in Equation  (\ref{KAP_DEF}).
	Besides, with the definition of the Green function in Equation (\ref{GRE_FUN}), one can find
	\begin{align}
		\Delta \mathcal G (\bm\theta,\bm\theta') = \delta (\cos\theta - \cos\theta') \delta (\phi - \phi') - \frac{1}{4\pi} \equiv \delta (\bm\theta - \bm\theta') - \frac{1}{4\pi}\ .\label{LAP_GRE}
	\end{align}
	With it one can rewrite Equation (\ref{PSI_DEF}) to Equation  (\ref{POI_EQU}).

	\section{Hints on the Proofs of the Lemmas}\label{APP_B}
	Without loss of generality, we assume the target coordinate system $K_T$ is the global one (i.e.,  $(\theta^{\sss (T)}, \phi^{\sss (T)}) = (\theta, \phi )$). Suppose the only point source is set on point$ (\theta_{\sss S},0)$, then for any point $(\theta,\phi)$ with  $\theta>\theta_{\sss S}  $ on the sphere, the potential is given by
	\begin{align}
		\psi &= \log \frac{1-\cos\theta \cos\theta_{\sss S} - \sin\theta\cos\phi \sin\theta_{\sss S}}{2}\\
		&= \log\left ( \frac{1}{4} (1+\cos\theta_{\sss S}) (1-\cos\theta) + \frac{1}{4} (1-\cos\theta_{\sss S})  (1+\cos\theta) - \frac{1}{2} \sin\theta_{\sss S} \sin\theta \cos\phi \right)  \\
		&= \log \left\{\cos^2 \frac{\theta_{\sss S}}{2}\sin^2 \frac{\theta}{2}\left[ \left ( 1 - \frac{\tan (\theta_{\sss S}/2)}{\tan (\theta/2) }\cos\phi\right)^2 + 
		\left (\frac{\tan (\theta_{\sss S}/2)}{\tan (\theta/2) } \sin\phi \right)^2 \right] \right\} \\
		&= 2 \log \cos\frac{\theta_{\sss S} }{2} + \log (1 - \cos\theta) - \log2 + 2 \Re \log\left ( 1 - \frac{\tan (\theta_{\sss S}/2)}{\tan (\theta/2) e^{i\phi} }\right) \\
		& = 2 \log \cos\frac{\theta_{\sss S} }{2} + \log (1 - \cos\theta) - \log2 - 2 \Re \sum_{l=1}^\infty \frac{1}{l} \left ( \frac{\tan (\theta_{\sss S}/2)}{\tan (\theta/2) e^{i\phi} } \right)^l\ .
	\end{align}
	If the point source is set at $ (\theta_{\sss S},\phi_{\sss S}) $, one only needs to rotate around the z-axis by $\phi_{\sss S}$, then repeat the calculation above.
	The rotation only produces a phase $e^{il\phi_{\sss S}}$ to the multipole with order $l$ so we get 
	
	\begin{align}
		\psi = 2 \log \cos\frac{\theta_{\sss S} }{2} + \log (1 - \cos\theta) - \log2 - 2 \Re \sum_{l=1}^\infty \frac{1}{l} \frac{z^l(\theta_{\sss S},\phi_{\sss S})}{z^l (\theta,\phi)}\ .
	\end{align}
	Note the local expansion of this potential can be obtained by exchange $(\theta_{\sss S},\phi_{\sss S})$ and $(\theta,\phi)$.
	For a set of particles on the sphere, by adding the contributions from all point sources, Expansion (\ref{LEMMA1:EXP}) in Lemma \ref{LEMMA1} is obtained. 
	
	The potential generated by a dipole at a point $(\theta_{\sss S},\phi_{\sss S})$ is given as
	\begin{align}
		\frac{1}{z (\theta^{\sss (S)},\phi^{\sss (S)})}  & =\tan\frac{\theta_{\sss S}}{2}
		+ \frac{2}{\sin\theta_{\sss S}}       \frac{z(\theta_{\sss S},\phi_{\sss S})}{z(\theta,\phi ) - z (\theta_{\sss S}, \phi_{\sss S})}\label{BASE_EXP1}\\
		&= \tan\frac{\theta_{\sss S}}{2} +
		\frac{2}{\sin\theta_{\sss S}} \sum_{l=1}^\infty \frac{z^l (\theta_{\sss S},\phi_{\sss S}) }{z^l(\theta,\phi) } \quad \text{for}\quad  \theta > \theta_{\sss S} \label{BASE_EXP2}\\
		&= - \cot\frac{\theta_{\sss S}}{2} -
		\frac{2}{\sin\theta_{\sss S}} \sum_{l=1}^\infty \frac{z^l(\theta,\phi)}{z^l(\theta_{\sss S},\phi_{\sss S})}  \quad \text{for} \quad \theta < \theta_{\sss S}\label{BASE_EXP3}
	\end{align}
	where $ (\theta^{\sss (S)}, \phi^{\sss (S)})$ is the coordinate in $K_{\sss S}$.
	Taking reciprocal on both sides of Equation  (\ref{BASE_EXP1}) gives
	\begin{align}
		z (\theta^{\sss (S)},\phi^{\sss (S)}) 
		&= - \tan\frac{\theta_{\sss S}}{2}  + \frac{2}{\sin \theta_{\sss S}} \frac{z(\theta,\phi)}{z(\theta,\phi) + z(\theta_{\sss S},\phi_{\sss S}) \cot^2 (\theta_{\sss S}/2)}\nonumber \\
		&=  - \tan\frac{\theta_{\sss S}}{2}  - \frac{2}{\sin \theta_{\sss S}}  \sum_{l=1}^{\infty} (-1)^l \tan^{2l}\left( \frac{\theta_{\sss S}}{2} \right) \frac{z^l(\theta,\phi)}{z^l(\theta_{\sss S},\phi_{\sss S})}\ .
	\end{align}
	Lemma \ref{LEMMA2}-\ref{LEMMA4} follow from these identities by exponentiation, substituting, and summation.

	
	\bibliography{sample631}{}

\begin{thebibliography}{}
\expandafter\ifx\csname natexlab\endcsname\relax\def\natexlab#1{#1}\fi
\providecommand{\url}[1]{\href{#1}{#1}}
\providecommand{\dodoi}[1]{doi:~\href{http://doi.org/#1}{\nolinkurl{#1}}}
\providecommand{\doeprint}[1]{\href{http://ascl.net/#1}{\nolinkurl{http://ascl.net/#1}}}
\providecommand{\doarXiv}[1]{\href{https://arxiv.org/abs/#1}{\nolinkurl{https://arxiv.org/abs/#1}}}

\bibitem[{Amara {et~al.}(2006)Amara, Metcalf, Cox, \&
  Ostriker}]{amara2006simulations}
Amara, A., Metcalf, R.~B., Cox, T.~J., \& Ostriker, J.~P. 2006, Monthly Notices
  of the Royal Astronomical Society, 367, 1367

\bibitem[{Becker(2013)}]{becker2013calclens}
Becker, M.~R. 2013, Monthly Notices of the Royal Astronomical Society, 435, 115

\bibitem[{Breton \& Reverdy(2021)}]{breton2021magrathea}
Breton, M.-A., \& Reverdy, V. 2021, arXiv preprint arXiv:2111.08744

\bibitem[{Carrier {et~al.}(1988)Carrier, Greengard, \&
  Rokhlin}]{carrier1988fast}
Carrier, J., Greengard, L., \& Rokhlin, V. 1988, SIAM journal on scientific and
  statistical computing, 9, 669

\bibitem[{Cheng {et~al.}(1999)Cheng, Greengard, \& Rokhlin}]{cheng1999fast}
Cheng, H., Greengard, L., \& Rokhlin, V. 1999, Journal of computational
  physics, 155, 468

\bibitem[{Das \& Bode(2008)}]{das2008large}
Das, S., \& Bode, P. 2008, The Astrophysical Journal, 682, 1

\bibitem[{de~Putter \& Takada(2010)}]{de2010halo}
de~Putter, R., \& Takada, M. 2010, Physical Review D, 82, 103522

\bibitem[{Dehnen(2002)}]{dehnen2002hierarchical}
Dehnen, W. 2002, Journal of Computational Physics, 179, 27

\bibitem[{Dominik {et~al.}(2002)Dominik, Albrow, Beaulieu, Caldwell, DePoy,
  Gaudi, Gould, Greenhill, Hill, Kane, {et~al.}}]{dominik2002planet}
Dominik, M., Albrow, M., Beaulieu, J.-P., {et~al.} 2002, Planetary and Space
  Science, 50, 299

\bibitem[{Dutton \& Maccio(2014)}]{dutton2014cold}
Dutton, A.~A., \& Maccio, A.~V. 2014, Monthly Notices of the Royal Astronomical
  Society, 441, 3359

\bibitem[{Ethridge \& Greengard(2001)}]{ethridge2001new}
Ethridge, F., \& Greengard, L. 2001, SIAM Journal on Scientific Computing, 23,
  741

\bibitem[{Giocoli {et~al.}(2017)Giocoli, Di~Meo, Meneghetti, Jullo, de~la
  Torre, Moscardini, Baldi, Mazzotta, \& Metcalf}]{giocoli2017fast}
Giocoli, C., Di~Meo, S., Meneghetti, M., {et~al.} 2017, Monthly Notices of the
  Royal Astronomical Society, 470, 3574

\bibitem[{Gorski {et~al.}(2005)Gorski, Hivon, Banday, Wandelt, Hansen,
  Reinecke, \& Bartelmann}]{gorski2005healpix}
Gorski, K.~M., Hivon, E., Banday, A.~J., {et~al.} 2005, The Astrophysical
  Journal, 622, 759

\bibitem[{Greengard \& Rokhlin(1987)}]{greengard1987fast}
Greengard, L., \& Rokhlin, V. 1987, Journal of computational physics, 73, 325

\bibitem[{Hezaveh \& Holder(2011)}]{hezaveh2011effects}
Hezaveh, Y.~D., \& Holder, G.~P. 2011, The Astrophysical Journal, 734, 52

\bibitem[{Hilbert {et~al.}(2009)Hilbert, Hartlap, White, \&
  Schneider}]{hilbert2009ray}
Hilbert, S., Hartlap, J., White, S., \& Schneider, P. 2009, Astronomy \&
  Astrophysics, 499, 31

\bibitem[{Hildebrandt {et~al.}(2016)Hildebrandt, Viola, Heymans, Joudaki,
  Kuijken, Blake, Erben, Joachimi, Klaes, Miller, Morrison, Nakajima,
  Verdoes~Kleijn, Amon, Choi, Covone, de~Jong, Dvornik, Fenech~Conti, Grado,
  Harnois-Déraps, Herbonnet, Hoekstra, Köhlinger, McFarland, Mead, Merten,
  Napolitano, Peacock, Radovich, Schneider, Simon, Valentijn, van~den Busch,
  van Uitert, \& Van~Waerbeke}]{KiDs}
Hildebrandt, H., Viola, M., Heymans, C., {et~al.} 2016, Monthly Notices of the
  Royal Astronomical Society, 465, 1454, \dodoi{10.1093/mnras/stw2805}

\bibitem[{Hoekstra {et~al.}(2004)Hoekstra, Yee, \&
  Gladders}]{hoekstra2004properties}
Hoekstra, H., Yee, H.~K., \& Gladders, M.~D. 2004, The Astrophysical Journal,
  606, 67

\bibitem[{Jain \& Seljak(1997)}]{jain1997cosmological}
Jain, B., \& Seljak, U. 1997, The Astrophysical Journal, 484, 560

\bibitem[{Jain {et~al.}(2000)Jain, Seljak, \& White}]{jain2000ray}
Jain, B., Seljak, U., \& White, S. 2000, The Astrophysical Journal, 530, 547

\bibitem[{Legin {et~al.}(2021)Legin, Hezaveh, Levasseur, \&
  Wandelt}]{legin2021simulation}
Legin, R., Hezaveh, Y., Levasseur, L.~P., \& Wandelt, B. 2021, arXiv preprint
  arXiv:2112.05278

\bibitem[{Meneghetti {et~al.}(2008)Meneghetti, Melchior, Grazian, De~Lucia,
  Dolag, Bartelmann, Heymans, Moscardini, \&
  Radovich}]{meneghetti2008realistic}
Meneghetti, M., Melchior, P., Grazian, A., {et~al.} 2008, Astronomy \&
  Astrophysics, 482, 403

\bibitem[{Meneghetti {et~al.}(2022)Meneghetti, Ragagnin, Borgani, Calura,
  Despali, Giocoli, Granato, Grillo, Moscardini, Rasia,
  {et~al.}}]{meneghetti2022probability}
Meneghetti, M., Ragagnin, A., Borgani, S., {et~al.} 2022, arXiv preprint
  arXiv:2204.09065

\bibitem[{Metcalf \& Petkova(2014)}]{10.1093/mnras/stu1859}
Metcalf, R.~B., \& Petkova, M. 2014, Monthly Notices of the Royal Astronomical
  Society, 445, 1942, \dodoi{10.1093/mnras/stu1859}

\bibitem[{M{\"o}ller \& Blain(1998)}]{moller1998strong}
M{\"o}ller, O., \& Blain, A. 1998, Monthly Notices of the Royal Astronomical
  Society, 299, 845

\bibitem[{M{\"o}ller \& Blain(2001)}]{moller2001strong}
---. 2001, Monthly Notices of the Royal Astronomical Society, 327, 339

\bibitem[{Navarro {et~al.}(1997)Navarro, Frenk, \& White}]{Navarro_1997}
Navarro, J.~F., Frenk, C.~S., \& White, S. D.~M. 1997, The Astrophysical
  Journal, 490, 493, \dodoi{10.1086/304888}

\bibitem[{Press {et~al.}(2007)Press, Teukolsky, Vetterling, \&
  Flannery}]{press2007numerical}
Press, W.~H., Teukolsky, S.~A., Vetterling, W.~T., \& Flannery, B.~P. 2007,
  Numerical recipes 3rd edition: The art of scientific computing (Cambridge
  university press)

\bibitem[{Sonnenfeld \& Cautun(2021)}]{sonnenfeld2021statistical}
Sonnenfeld, A., \& Cautun, M. 2021, Astronomy \& Astrophysics, 651, A18

\bibitem[{Takahashi {et~al.}(2017)Takahashi, Hamana, Shirasaki, Namikawa,
  Nishimichi, Osato, \& Shiroyama}]{takahashi2017full}
Takahashi, R., Hamana, T., Shirasaki, M., {et~al.} 2017, The Astrophysical
  Journal, 850, 24

\bibitem[{Taruya {et~al.}(2002)Taruya, Takada, Hamana, Kayo, \&
  Futamase}]{taruya2002lognormal}
Taruya, A., Takada, M., Hamana, T., Kayo, I., \& Futamase, T. 2002, The
  Astrophysical Journal, 571, 638

\bibitem[{Teyssier {et~al.}(2009)Teyssier, Pires, Prunet, Aubert, Pichon,
  Amara, Benabed, Colombi, Refregier, \& Starck}]{teyssier2009full}
Teyssier, R., Pires, S., Prunet, S., {et~al.} 2009, Astronomy \& Astrophysics,
  497, 335

\bibitem[{Thompson {et~al.}(2010)Thompson, Fluke, Barnes, \&
  Barsdell}]{thompson2010teraflop}
Thompson, A.~C., Fluke, C.~J., Barnes, D.~G., \& Barsdell, B.~R. 2010, New
  Astronomy, 15, 16

\bibitem[{Vale \& White(2003)}]{vale2003simulating}
Vale, C., \& White, M. 2003, The Astrophysical Journal, 592, 699

\bibitem[{Wambsganss(1999)}]{wambsganss1999gravitational}
Wambsganss, J. 1999, Journal of Computational and Applied Mathematics, 109, 353

\bibitem[{Wambsganss {et~al.}(1998)Wambsganss, Cen, \&
  Ostriker}]{wambsganss1998testing}
Wambsganss, J., Cen, R., \& Ostriker, J.~P. 1998, The Astrophysical Journal,
  494, 29

\bibitem[{Wei {et~al.}(2018)Wei, Li, Kang, Luo, Xia, Wang, Yang, Wang, Jing,
  Mo, {et~al.}}]{wei2018full}
Wei, C., Li, G., Kang, X., {et~al.} 2018, The Astrophysical Journal, 853, 25

\bibitem[{Wright \& Brainerd(2000)}]{wright2000gravitational}
Wright, C.~O., \& Brainerd, T.~G. 2000, The Astrophysical Journal, 534, 34

\bibitem[{Xu \& Jing(2021)}]{xu2021accurate}
Xu, K., \& Jing, Y. 2021, The Astrophysical Journal, 915, 75

\bibitem[{Ying {et~al.}(2004)Ying, Biros, \& Zorin}]{ying2004kernel}
Ying, L., Biros, G., \& Zorin, D. 2004, Journal of Computational Physics, 196,
  591

\end{thebibliography}
	\bibliographystyle{aasjournal}
	
	

\end{document}